\documentclass{article}

\usepackage{graphicx}
\usepackage{amsmath}
\usepackage{amssymb}
\usepackage{float}
\usepackage{caption}
\usepackage{color}
\usepackage{soul}
\usepackage{hyperref}
\graphicspath{{../PNAS-PLUS/ThesisChapter/images/}{../PNAS-PLUS/ThesisChapter/supplement/}}

\usepackage{upgreek}
\usepackage{enumitem}

\newcommand{\dund}[1]{\smash{\underline{#1}}}
\begin{document}
\title{
Cell Shape and Durotaxis Follow from Mechanical Cell-Substrate Reciprocity and Focal Adhesion Dynamics: A Unifying Mathematical Model}



\author{Elisabeth G. Rens (1,2,3) and Roeland M.H. Merks (1,2,4)}

\date{\small(1) Scientific Computing, CWI, Science Park 123, 1098 XG Amsterdam, The Netherlands \\
 (2) Mathematical Institute, Leiden University, Niels Bohrweg 1, 2333 CA Leiden, The Netherlands \\
 (3) Present address: Mathematics Department, University of British Columbia, Mathematics Road 1984, V6T 1Z2, Vancouver, BC, Canada, {\tt rens@math.ubc.ca}\\
 (4) Present address: Mathematical Institute and Institute for Biology, Leiden University, Niels Bohrweg 1, 2333 CA Leiden, The Netherlands, {\tt merksrmh@math.leidenuniv.nl}
 }

\maketitle

\begin{abstract}
Many animal cells change their shape depending on the stiffness of the substrate on which they are cultured: they assume small,  rounded shapes in soft ECMs, they elongate within stiffer ECMs, and flatten out on hard substrates. Cells tend to prefer stiffer parts of the substrate, a phenomenon known as durotaxis.  Such mechanosensitive responses to ECM mechanics are key to understanding the regulation of biological tissues by mechanical cues, as it occurs, e.g., during angiogenesis and the alignment of cells in muscles and tendons.  Although it is well established that the mechanical cell-ECM interactions are mediated by focal adhesions, the mechanosensitive molecular complexes linking the cytoskeleton to the substrate, it is poorly understood how the stiffness-dependent kinetics of the focal adhesions eventually produce the observed interdependence of substrate stiffness and cell shape and cell behavior. Here we show that the mechanosensitive behavior of single-focal adhesions, cell contractility and substrate adhesivity together suffice to explain the observed stiffness-dependent behavior of cells. We introduce a multiscale computational model that is based upon the following assumptions: (1) cells apply forces onto the substrate through FAs; (2) the FAs grow and stabilize due to these forces; (3) within a given time-interval, the force that the FAs experience is lower on soft substrates than on stiffer substrates due to the time it takes to reach mechanical equilibrium; and (4) smaller FAs are pulled from the substrate more easily than larger FAs. Our model combines the cellular Potts model for the cells with a finite-element model for the substrate, and describes each FA using differential equations. Together these assumptions provide a unifying model for cell spreading, cell elongation and durotaxis in response to substrate mechanics. 
\end{abstract}
Classification: Physical Sciences, Applied Mathematics 
\newpage

\section*{Significance statement}
Besides molecular signaling, mechanical cues coordinate cell behavior during embryonic development and wound healing; for example, in tendons cells and collagen fibers align to optimally support the forces that the tendon experiences. To this end, cells actively probe their environment and respond to its mechanics. Key building blocks for such active mechanosensing are the contractile actin cytoskeleton and the extracellular matrix (ECM), the matrix of fibrous proteins that glues cells together into tissues (e.g., collagen). Actin is linked to the ECM by structures called focal adhesions, which stabilize under force. Here we present a novel mathematical model that shows that these building blocks suffice to explain the cellular responses to ECM stiffness. The insights advance our understanding of cellular mechanobiology. 


\newpage

\section{Introduction}
Embryonic development, structural homeostasis and developmental diseases are driven by biochemical signals and biomechanical forces. By interacting with the extracellular matrix (ECM), a network of fibers and proteins that surrounds cells in tissues, cells can migrate and communicate with other cells, which contributes to tissue development. Mechanical interactions between cells and the ECM are crucial for the formation and function of tissues. By sensing and responding to physical forces in the ECM, cells change their shape and migrate to other locations. Here, we show a single, unifying mechanism that suffices to explain such ECM-mechanics induced cell shape changes and cell migration.

The shape of a wide range of mammalian cell types depends on the stiffness of the ECM. In vitro, cells cultured on top of soft, two-dimensional ECM substrates become relatively small and rounded, whereas on top of a stiffer ECM the cells assume elongated shapes. On ECM of high rigidity, like glass, cells spread out and flatten. This behavior has been observed for a wide range of cell types, including endothelial cells \cite{Califano2010}, fibroblasts \cite{Pelham:1997uc,Ghibaudo2008}, smooth muscle cells \cite{Engler2004}, and osteogenic cells \cite{Mullen2015}). Secondly, cells tends to migrate towards stiffer parts of the ECM, a phenomenon known as durotaxis. Such behavior also occurs for a wide range of mammalian cell types, including fibroblasts~\cite{Lo:2000cj}, vascular smooth muscle cells~\cite{Isenberg2009} and mesenchymal stem cells \cite{Vincent2013}. 
 
It is still poorly understood what molecular mechanisms regulate such cellular response to ECM stiffness \cite{Jansen2015}. Cells are able to sense matrix stiffness through focal adhesions (FAs), multi-molecular complexes consisting of integrin molecules that mediate cell-ECM binding and force transmission, and an estimated further 100 to 200 protein species that strongly or more loosely associate with focal adhesions \cite{Bershadsky2003,ZaidelBar:2007hr,WinogradKatz:2014dg}. Among these are vinculin and talin, which bind integrin to actin stress fibers. 
Manipulations of FA assembly affects cell spreading and motility. For instance, a lack of vinculin directly decreases cell spreading \cite{Ezzel1997} through FA stabilization, even when actin stress fibers are not affected. Generally, changes in cell polarization were associated with altered FA response to substrate rigidity (through gene knock-downs) \cite{Prager2011}. FA assembly and disassembly has also been associated to cell migration and orientation \cite{Broussard2008,Plotnikov:2012cp,Chen2013,Kim2013} in response to the ECM. So, the mechanosensitive growth of FAs is key to our understanding of how cells respond to ECM stiffness.

Focal adhesions dynamically assemble and disassemble, where the disassembly rate is highest on soft ECMs and is lower on stiffer ECMs \cite{Pelham:1997uc} leading to FA-stabilization. This mechanosensitivity of FA dynamics is regulated by FA-proteins, such as talin and p130Cas. These proteins change conformation in response to mechanical force \cite{Bershadsky2003,Jansen2015}. For instance, stretching the structural protein talin reveales vinculin binding sites, allowing additional vinculin to bind to focal adhesions \cite{Rio2009} and stabilize the FA \cite{Gallant2005}. Furthermore, integrins such as $\alpha_5 \beta_1$, behave as so-called ``catch-bonds'' \cite{Kong2009}: bonds whose lifetime increase under force \cite{Dembo1988}. However, how different mechanosensitive proteins in FAs and cytoskeletal forces work together to regulate cell spreading, cell shape, and durotaxis is still to be elucidated.

Previous mathematical models have proposed various explanations for the mechanosensitive behavior of cells. A central idea in these explanations is dynamic reciprocity \cite{Bissell:1982wb}:  the cell pulls on the ECM to probe its mechanical response. The mechanical resistance of the matrix can either lead the cell to pull more strongly, or it can change the stability of the focal adhesions. 
Ni et al. \cite{Ni2007} assume that cells exert stronger traction forces on stiffer ECMs than on more soft ECMs. The resulting balance between intracellular and extracellular stresses leads the cell to deform; the deformation is attenuated by an interfacial energy due to cell-ECM adhesion. This model predicts that cells deform only if the cell and substrate stiffness are roughly equally stiff, whereas cell spreading forces are not included in the model. To show how stresses in the ECM can affect cell polarization, Shenoy et al. \cite{Shenoy2015} argue that deformation of cells can lead to stress-induced cell polarization of contractility. They show that stress-induced recruitment of myosin to the actin cytoskeleton can locally increases contractility, leading to further intracellular stress. The resulting positive feedback loop can polarize cells, contributing to cell elongation and durotaxis. 
In our own work, we have proposed that cells spread and elongate because of a positive feedback loop between protrusion forces and strain stiffening of the matrix \cite{Oers:2014pcb}, a mechanism motivated by the  stabilization of focal adhesions on stiff matrices. 
Similarly, previous models have explained durotaxis. The prevalent idea in these mathematical models is increased stabilization of focal adhesions at the stiff side of the cell, which in turn drives intracellular dynamics, leading to increased traction force \cite{Dokukina2010},  modified stress fiber dynamics \cite{Lazopoulos,Harland2011}, a bias in velocity \cite{Stefanoni2012}, viscous forces and cell stiffening \cite{Aubry2015}, motor protein recruitment \cite{Shenoy2015}, membrane tension \cite{Allena2016}, enhanced persistence time in stiffer matrices \cite{Novikova2017}, or cell polarization \cite{Yu2017,Kim2018}.

Some models integrated explicit kinetics of focal adhesions \cite{DESHPANDE20081484} in cell-based models to study how force dependent FA assembly regulates cell spreading \cite{Ronan2014,Vernerey2014}. In these models the size of the focal adhesion (i.e., the number of cell-substrate adhesions within it) determines how much force cells apply on the matrix. These models assumed that on stiff substrates, the experienced stress promotes stress fiber formation in the cell, allowing it to apply more force on the focal adhesions. As a result, the focal adhesions grow and the cell can apply even more force, stabilizing the stress fibers even more, etc. In both these models, such a feedback between force and focal adhesions makes the cell spread out. Another hybrid cell and focal adhesion model (that integrated the focal adhesion growth model in ref. \cite{Besser2006}) where the size of the focal adhesions did not affect the magnitude of cell forces, could not explain increased cell spreading on stiff matrices \cite{Stolarska2017}. These results suggest that cell spreading is regulated by a feedback between focal adhesions and traction force. These models could however not predict how cells elongate as a function of substrate stiffness. 

With a hybrid cell and focal adhesion model we propose a focal adhesion mechanism that unifyingly explains three ECM stiffness-dependent cell behaviors: cell spreading, cell elongation, and durotaxis. The model is based on the following assumptions: (1) focal adhesions are discrete clusters of integrin-ECM bonds; (2) new bonds are added to the FAs at a constant rate; (3) the unbinding rate is suppressed by the tension in the FA, which is due to pulling of stress fibers \cite{Novikova2013}; (4) on soft ECMs it takes more time for the tension in the FA to build up to its maximum value, than on stiff ECMs \cite{Schwarz2006}; therefore, on average, the unbinding rate in FAs is higher on soft ECMs than on stiff ECMs. (5) As a result, FAs grow larger on stiff ECMs than on softer ECMs; (6) thus the FAs detach less easily from the ECM on stiff matrices than on softer ones. (7) Planar stress reinforces FAs due to recruitment of stabilizing proteins such as vinculin. 

We show that, apart from the substrate-stiffness-dependent cell traction forces proposed in the previous works, an alternative explanation for cell spreading on stiff substrates is that cells build up their forces with a \textit{faster rate} on stiff substrates. Interestingly, in our model, a feedback between focal adhesions and force magnitude, regulates cell elongation (and not cell spreading itself). The model shows that the range of stiffness on which cells elongate, depends on the velocity of myosin motor proteins. Finally, simulated cells exhibit durotaxis and consistent with experimental observations, the durotaxis speed increases with the slope of the stiffness gradient \cite{Isenberg2009,Vincent2013} .


\section*{Results}

\begin{figure*}
\centering
\includegraphics[width=\textwidth]{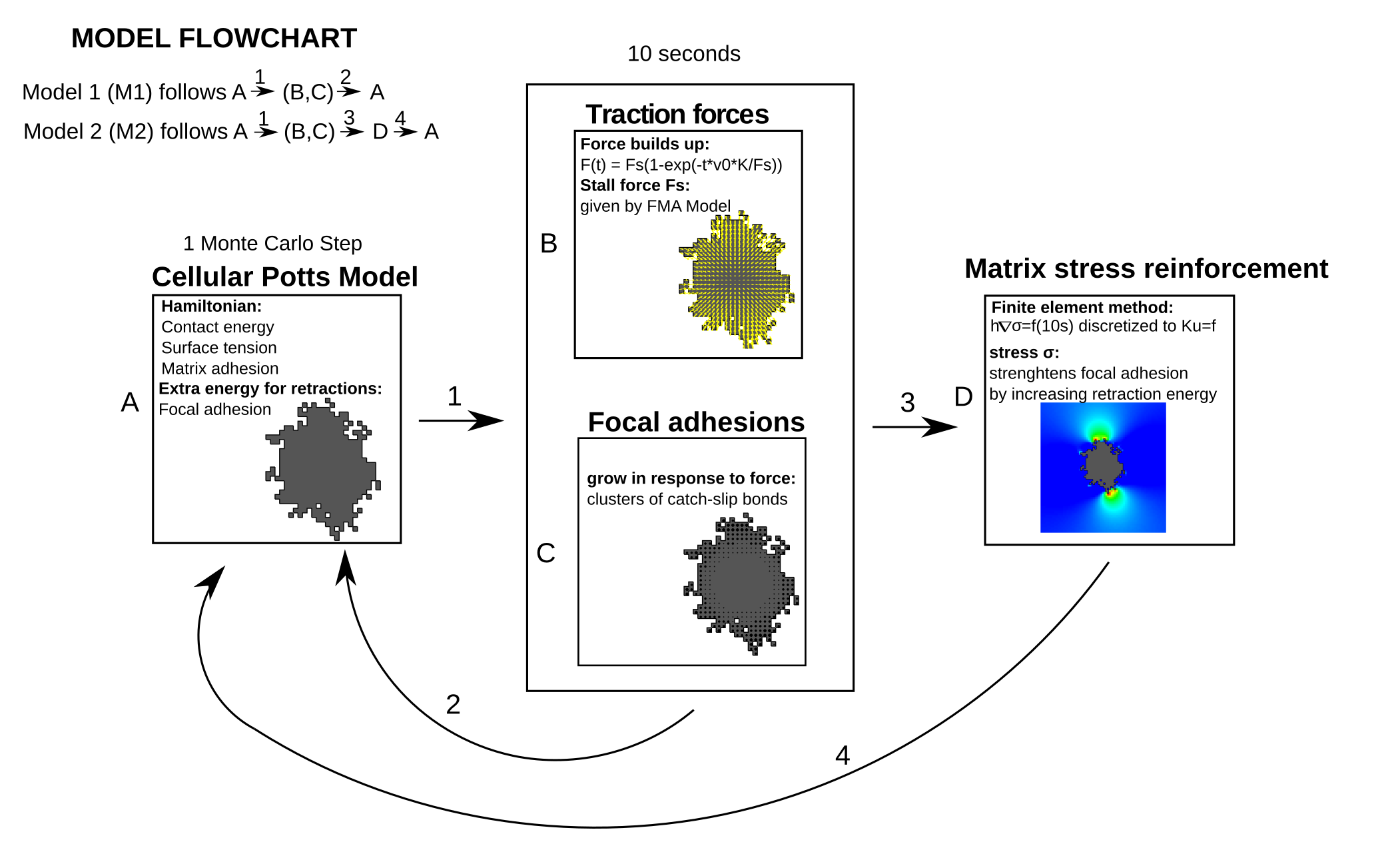}
\caption{Flowchart of the multiscale CPM. (A) CPM calculates cell shapes in response to focal adhesions and substrate stresses; (B) calculation of cellular traction forces based on cell shape and force build-up dynamics; (C) focal adhesion grow according to dynamics of catch-slip bond clusters; and (D) calculation of substrate stresses due to cellular traction forces.}
\label{fig:fig1}
\end{figure*}

\noindent\textbf{Model development}: We based our model on a recently developed hybrid Cellular Potts - Finite Element framework that has been tested on collective cell behavior driven by ECM forces \cite{Oers:2014pcb,Rens2017}. We extended and adapted this model to include explicit descriptions of focal adhesion dynamics. With this model, we propose that focal adhesion dynamics can explain cell spreading, cell elongation and durotaxis in response to substrate stiffness. Figure~\ref{fig:fig1} gives an overview of the model, showing the flow and feedback between the cell, its focal adhesions and the elastic substrate it adheres to.  Here, we will give a brief explanation of the model. We refer to the methods section for more details.

\noindent\textbf{Cells}: A cell is described as a collection of discrete lattice sites in a cellular Potts model (CPM), see Figure~\ref{fig:fig1}A. Cells in the CPM change shape by iteratively making extensions and retractions, modeling the formation and break down of adhesions with the substrate during so-called Monte Carlo time steps (MCS). We assume that retractions from the substrate are less likely at sites with large focal adhesions.

\noindent\textbf{Cell traction forces}: The cell applies a contractile force upon focal adhesions that adhere to the matrix. We use the shape of the cell to calculate the contractile force based on a First Moment of Area (FMA) model \cite{Lemmon:2010ju}, see Figure~\ref{fig:fig1}B. This model assumes that the cell acts as a single contractile unit, so that the resulting forces are pointed towards the center of mass and forces are proportional to the distance to the center of mass. To describe the force dynamics in time, we adopt a model of Schwarz \textit{et al.} \cite{Schwarz2006}: $$F(t) = F_s(1-\exp({-t \cdot \frac{v_0 K}{F_s}})),$$ where $v_0$ is the free velocity of the motor proteins and $K$ the substrate stiffness and $F_s$ is the stall force. This model assumes that forces build-up towards the stall force (here, given by the FMA model) with a rate proportional to the velocity of the myosin motors (that are responsible for force generation) and the matrix stiffness. So, the more compliant the substrate is, the longer it takes for a cell to build up this force. 

\noindent\textbf{Focal adhesions}: We model focal adhesions as clusters of catch-slip bonds, as proposed by Novikova and Storm \cite{Novikova2013}. This model is based on experimental data of one single $\alpha5-\beta1$ integrin, of which the degradation rate of bound integrins decreases with force. Novikova and Storm then extended their model to describe the dynamics of a cluster of integrins, one focal adhesion. At each site of our 2D CPM, we implement the dynamics of one such focal adhesion (see Figure~\ref{fig:fig1}C).

\noindent\textbf{Simulation}: A simulation proceeds as follows. We initiate cells in the CPM (part A of our model). Then, in part B of the model, we let those forces build up in time for $t_\mathrm{FA}$ seconds. At the same time, part C of the model is executed. So as the forces build up, we let the integrin clusters grow simultaneously. After these $t_\mathrm{FA}$ seconds, we let the cells move, i.e. perform one timestep in the CPM. After one timestep of the CPM, we again let the forces build up and the integrin clusters grow for $t_\mathrm{FA}$, and so forth. Parameter values were, if possible, chosen based on literature and if not, chosen arbitrarily. For the arbitrarily chosen parameters, we performed parameter variations to study their effects (see Supplementary Material), which showed that deviations from the default values do not qualitatively change our results. The default parameter values are given in Supplementary Table S1.

\noindent\textbf{`Minimal' model M1}: In a minimal version of our model (called model M1), we only follow the loop with arrows 1 and 2 in the flowchart, thus excluding the feedback with matrix stress as depicted in Figure~\ref{fig:fig1}D (which is not described here, but will be describe in a later section where it will be used for the first time). We start out with model M1, to study how this minimal model translates to cell spreading. 

\subsection*{Catch-bond cluster dynamics suffices to predict cell area as a function of substrate stiffness} 
\begin{figure}
\centering
\includegraphics[width=\textwidth]{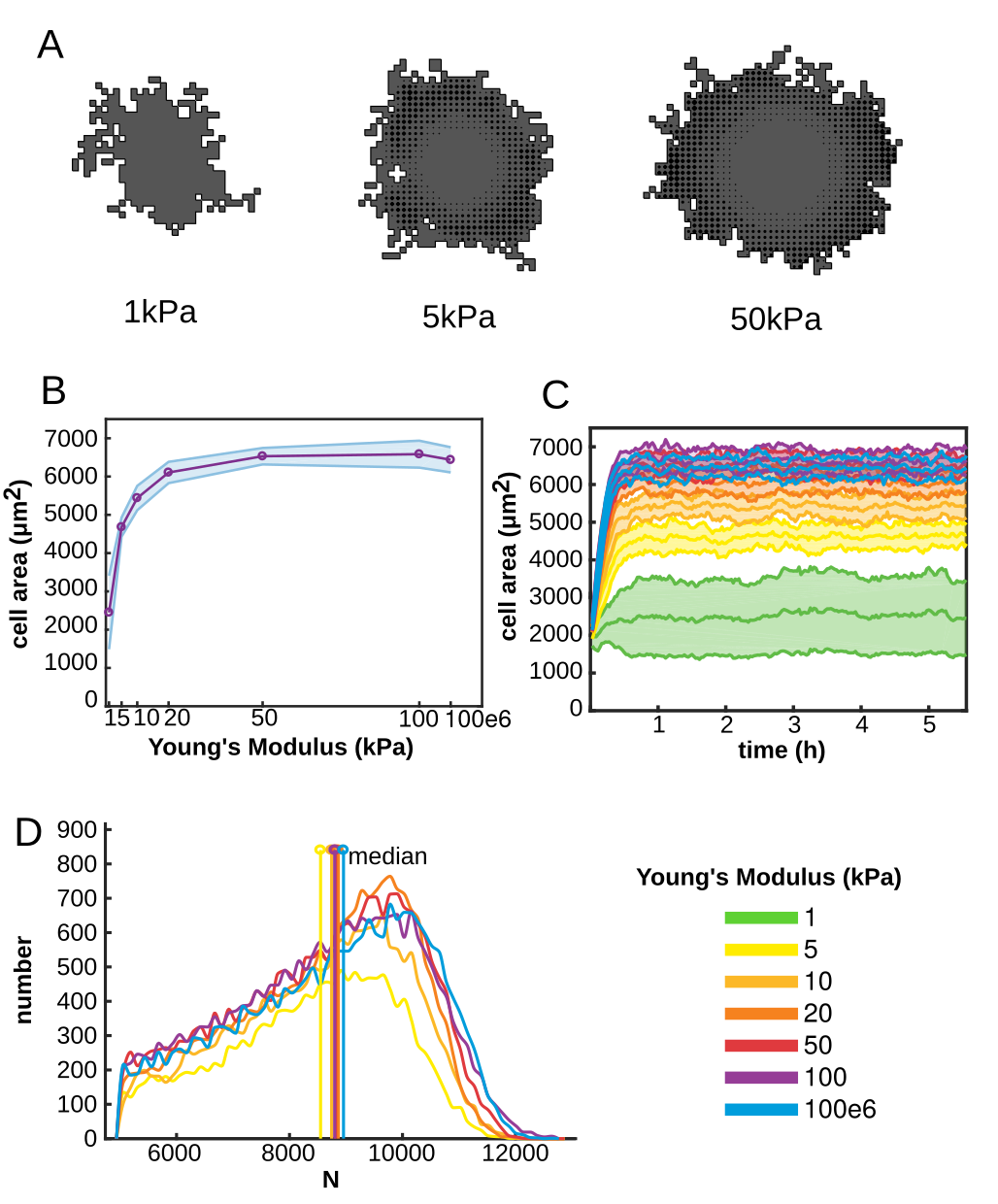}
\caption{Model M1 predictions: Cell area increases with increasing substrate stiffness. (A) Example configurations of cells at 2000 MCS on substrates of 1,50 and 50 kPa; (B) Cell area as a function of substrate stiffness, shaded regions: standard deviations over 25 simulations; (C) Timeseries of cell area, shaded regions: standard deviations over 25 simulations; and (D) distribution of N, the number of integrin bonds per cluster, all clusters at 2000 MCS from 25 simulations were pooled. We indicate the median. Color coding (C and D): See legend next to (D).}
\label{fig:fig2}
\end{figure}

In this section, we show with model M1, that catch-bond dynamics of integrin clusters suffices to explain cell spreading on elastic substrates. We set up the simulations as described above and varied the substrate stiffness. Figure~\ref{fig:fig2}A shows the response of cells on substrate of size 500 $\upmu$m by 500 $\upmu$m with a stiffness of $1\;\mathrm{kPa}$, $5\;\mathrm{kPa}$, and $50\;\mathrm{kPa}$ after 2000 MCS ($\approx$ 5.5 h) (see also Video S1). On the most soft substrate of $1\;\mathrm{kPa}$, focal adhesions do not grow and the cell does not spread. On a slightly stiffer substrate of 5 kPa, focal adhesions have grown and the cell has significantly increased in size. Increasing the substrate further also increases the cell area, although from 50 kPa the cell does not seem to change in size. On stiffer substrates, there are more larger focal adhesions visible and they seem to accumulate more around the cell membrane, as shown in the two insets in Figure~\ref{fig:fig2}A. 

Figure~\ref{fig:fig2}B plots the cell area as a function of substrate stiffness. Cell area increases from around 2500 $\upmu$m$^2$ on the softest substrate and plateaus at around 6500 $\upmu$m$^2$ at a stiffness of 50 kPa. Thus, the cell area has increased more than 2.5 fold on the stiffest substrate compared to the softest substrates. This factor is consistent with experimental observations \cite{Califano2010,Balcioglu1316,Asanoe13281}. We also investigated if the model could qualitatively predict spreading dynamics. Figure~\ref{fig:fig2}C plots the cell area as a function of time. The cells quickly increase in size and and reach their final size after 30 to 60 minutes. Experimental curves of cell area versus time follow a similar trend \cite{ReinhartKing:2005dq,Yeung2005}. 

We also investigated the distribution of the integrin cluster sizes. Figure~\ref{fig:fig2}D plots the distribution of the cluster sizes and the median cluster size (average cluster size is roughly the same). The median cluster size and variance are unaffected by substrate stiffness, in contrast with experimental observations \cite{Paszek:2005eh}. We performed a more detailed analysis of the distribution of integrin clusters and describe two observations. 1) On stiffer substrates, there are more larger clusters. For instance, the percentage of focal adhesions with $N>10000$ is 20\% on 50000 kPa, 15\% on 10 kPa and 10\% on 5 kPa. 2) On stiffer substrates, large focal adhesions are found at the cell boundary (Figure~\ref{fig:fig2}A for 50 kPa and Supplementary Figure~{1}). On soft substrates, large focal adhesions are found further away from the cell center, where forces had time to build up because in the bulk of the cell no retractions take place (Figure~\ref{fig:fig2}A for 5 kPa and Supplementary Figure~{1}). 

All in all, the results presented in this section suggest that the catch-slip bond dynamics of single integrins within focal adhesions suffice to predict cell area and spreading dynamics from substrates stiffness. Our model shows that cells can spread due to the intertwined dynamics of force build-up, focal adhesion growth and cell-matrix adhesion. On soft substrates, forces build up slowly, so there is not enough time for a focal adhesion to grow to strongly adhere the cell to the matrix. So, the cell will continuously make extensions and retractions. In contrast, on stiff substrates, forces build up fast and focal adhesions are able to grow and extensions have a long lifetime, allowing the cell to spread.

\subsection*{Focal Adhesion strengthening due to matrix stress induces cell elongation}
After having captured, at a qualitative level, the rate of spreading as a function of substrate stiffness, we now set out to explain the ability of mammalian cells to elongate on stiff enough substrates. Because the first version of the model (M1) could not yet explain cell elongation, we aimed to find an additional focal adhesion mechanism that can explain cell elongation.

Since cell traction forces are transferred to the matrix through the integrins, stresses develop in the matrix. Such stresses have been observed to affect focal adhesion assembly \cite{Chen2013}. We therefore hypothesized that such a feedback may explain cell shape changes. 

\noindent\textbf{`Extended' model M2}: We extended our model to Model M2, that includes a finite element model (FEM) to calculate the matrix stress (Figure~\ref{fig:fig1}D) as a result of the cell traction forces \cite{Oers:2014pcb,Rens2017}. So, model M2 follows arrows 1, 3 and 4 in Figure~\ref{fig:fig1}D. We assume that matrix stresses reinforces cell-matrix adhesions. We model focal adhesion strengthening by reducing the probability of retractions from the matrix due to matrix stress, \textit{i.e.} we multiply the energy it takes for a cell to make a retraction with $$ 1+p \frac{g(\dund{\sigma}(\vec{x}'))}{\sigma_h+g(\dund{\sigma}(\vec{x}'))}.$$ Here, parameter $p$ regulates the strengthening and $\sigma_h$ its saturation and $g(\dund{\sigma}(\vec{x}'))$ denotes the hydrostatic stress on the lattice site of retraction. Such a strengthening due to matrix stress can have various molecular origins. We hypothesize that this strengthening is due to stretching of the structural protein talin exposes binding sites for vinculin, which binds to the cytoskeleton and thus strengthens the actin-integrin linkage \cite{Rio2009,Gallant2005}. 

\begin{figure}
\centering
\includegraphics[width=\textwidth]{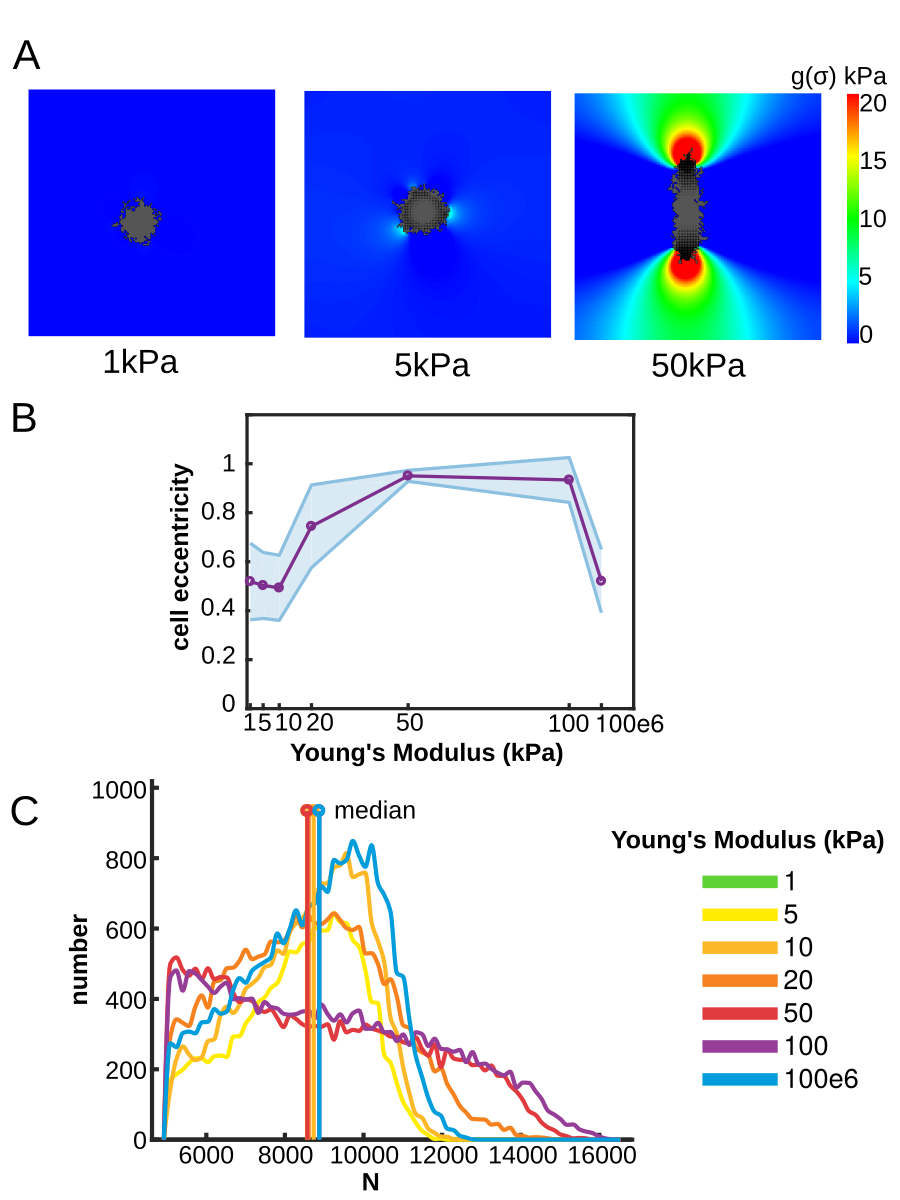}
\caption{Model M2 predictions: Cells elongate on substrates of intermediate stiffness. (A) Example configurations of cells at 2000 MCS on substrates of 1,50 and 50 kPa. Colors: hydrostatic stress; (B) Cell eccentricity as a function of substrate stiffness, shaded regions: standard deviations over 25 simulations; (C) distribution of N, the number of integrin bonds per cluster, all focal adhesion at 2000 MCS from 25 simulations were pooled. We indicate the median. Color coding (C): See legend next to (C).}
\label{fig:fig3}
\end{figure}

Figure~\ref{fig:fig3}A shows representative configurations of cells after running model M2 (see also Video S2). Similar to model M1, on the most soft substrate (1 kPa), the cell stays small and round. However, for stiffer substrates, such as 50kPa, the cell elongates. To quantify cell elongation, we measured the eccentricity of cells as $\sqrt(1-\frac{b^2}{a^2})$ with $a$ and $b$ the lengths of the cell's major and minor semi-axes, calculated as the eigenvalues of the inertia tensor. Figure~\ref{fig:fig3}B shows that cells start to slightly elongate on substrates with a stiffness of 10 kPa/20 kPa. On stiffer matrices, 50 kPa and 100 kPa, cells are very much polarized in shape and large focal adhesions have grown at the tips of the cell. On the very rigid substrate, the cell is more circular again. So, the eccentricity of cells has a biphasic dependence on substrate stiffness. We also again quantified the distribution of the integrin cluster sizes. Figure~\ref{fig:fig3}C shows the distribution of the cluster sizes for the different elastic substrates. The median (which is again roughly the same as the mean) cluster size does not vary much between substrate stiffness. The shape of the distributions, however, are more flat and with higher standard deviation on the substrates where cells have elongated compared to round cells. The kurtosis is around 2.0 for elongated cells, compared to a kurtosis around 2.3 for round cells. The standard deviation is around 2700 for elongated cells, compared to around 1500 for round cells. This more flat distribution of focal adhesion sizes can be explained as follows. An elongated shape results in large traction force at the tip of the cells, such that the focal adhesions grow larger in size at these tips, while at the sides of the cell, the forces are much smaller and focal adhesion stay small there.

The model explains the process of cell elongation as follows. On sufficiently stiff matrices, the cell initially starts to spread. The cell continuously makes random protrusions, allowing the cell shape to become slightly anisotropic. Around these cell protrusions, matrix stresses develop, which strengthens cell-matrix adhesion in this region. So, the cell can continue to build up forces, allowing the focal adhesion to grow larger. In contrast, at site of lower matrix stress, focal adhesions are more likely to disassemble. At protruding sites, cell traction forces increase due to an increased distance from the cell centroid. This results in a breaking of symmetry and the cell starts to elongate due to a positive feedback loop of force build-up, focal adhesion growth and matrix stress induced adhesion strengthening. On soft matrices, matrix stresses are not high enough to initiate a symmetry breaking. On the most rigid surface, matrix stresses are too high, allowing focal adhesions to strengthen equally well such that no symmetry breaking can occur. So, cell elongation is only possible on substrates with an optimal rigidity. Similar dynamics of cell spreading followed by a symmetry breaking has also been observed experimentally \cite{Prager2011}. Note that the spindle-like shapes that cells obtain in our model resembles those observed \textit{in vitro} \cite{Califano:2010dp}.

Some parameters, such as $p$ and $\sigma_h$ were chosen arbitrarily. So, we tested the sensitivity of our model M2 to these parameters. Increasing $p$ which regulates the extent of focal adhesion strengthening by matrix stress, enables cells to start elongating on softer matrices and also induces cell elongation on the most rigid surface (Supplementary Figure~{2}). Variations in $\sigma_h$, which regulates the saturation of stretch exposed binding sites for vinculin does not greatly affect model behavior (Supplementary Figure~{3}). Other parameters might be cell type specific, such as the lifetime of protrusions $t_\mathrm{FA}$ (Supplementary Figure~{4}), extent of random motility $T$ (Supplementary Figure~{5}) and the magnitude of traction forces $\mu$ (Supplementary Figure~{6}). The qualitative behavior is conserved for variations of these parameters, but all parameters affect the range of substrate stiffness on which the cell can elongate.

\noindent\textbf{Second version of `extended' model M2}: 
Since previous models proposed that cell spreading is regulated by a feedback between focal adhesions and forces, we also investigate what happens if focal adhesions start to apply more force when subject to matrix stresses. So, instead of letting matrix stresses strenghten focal adhesions by increasing the detachment energy, we let matrix stress strengthen the focal adhesions by locally increasing cell traction forces. To model this, we assume that the stall force increases as a function of matrix stress, \textit{i.e.} $\vec{F}_s = \vec{F}_s \cdot \left ( 1+p \frac{g(\dund{\sigma}(\vec{x}'))}{\sigma_h+g(\dund{\sigma}(\vec{x}'))} \right )$. This other feedback mechanism gives similar results to Figure~\ref{fig:fig3} (see Supplementary Figure~{7} and Video S3 for the results of Model M2-version 2). Such a mechanism can have various molecular origins. For instance, addition of vinculin through talin stretching can induce increased traction forces \cite{Dumbauld2013}. Stretching forces also induces $\alpha$-smooth muscle actin recruitment to stress fibers \cite{Pittet2006}, and myosin motor binding \cite{Uyeda2011}. 

In conclusion, our model suggests that by applying a force on the matrix, cells develop an anistropic matrix stress field that can induce a symmetry breaking of the cell by reinforcing focal adhesion sites. This allows a cell to elongate on substrates of intermediate stiffness. Such a matrix stress reinforcement can be  from various molecular origins, such as a matrix stress induced focal adhesion strengthening or increased traction forces.

\subsection*{Motor protein velocity changes stiffness regime on which cells elongate}

The regime of substrate stiffness on which cells spread and elongate varies per cell type. For instance, neutrophils do not respond to changes of substrate stiffness in the range of substrate stiffness where both fibroblasts and endothelial change in area and shape \cite{Yeung2005}. To try and understand why this is the case, we can vary cell related parameters in our model. One cell specific parameter is the velocity of the myosin motors, which regulates the speed of force build-up. Many cells express non-muscle myosin II, which exists in isoforms A,B and C \cite{Ma2012}. Other cell types also expresses myosin isoforms such as skeletal, cardiac and smooth muscle myosin \cite{Ma2012}. Different cell types may have different expression profiles of myosin isoforms \cite{Murakimi1993} and since the velocity of myosin motors varies among isoforms \cite{Norstrom2010,Kelley1996}, this may impact the response of cells to matrix stiffness. Using our model, we study how myosin motor velocity, $v_0$, can impact cell shape. We study a range from 10 nm/s, corresponding to non-muscle myosin II B \cite{Norstrom2010}) to 1000 nm/s, corresponding to muscle myosin \cite{Vogel2013}.

\begin{figure}
\centering
\includegraphics[width=\textwidth]{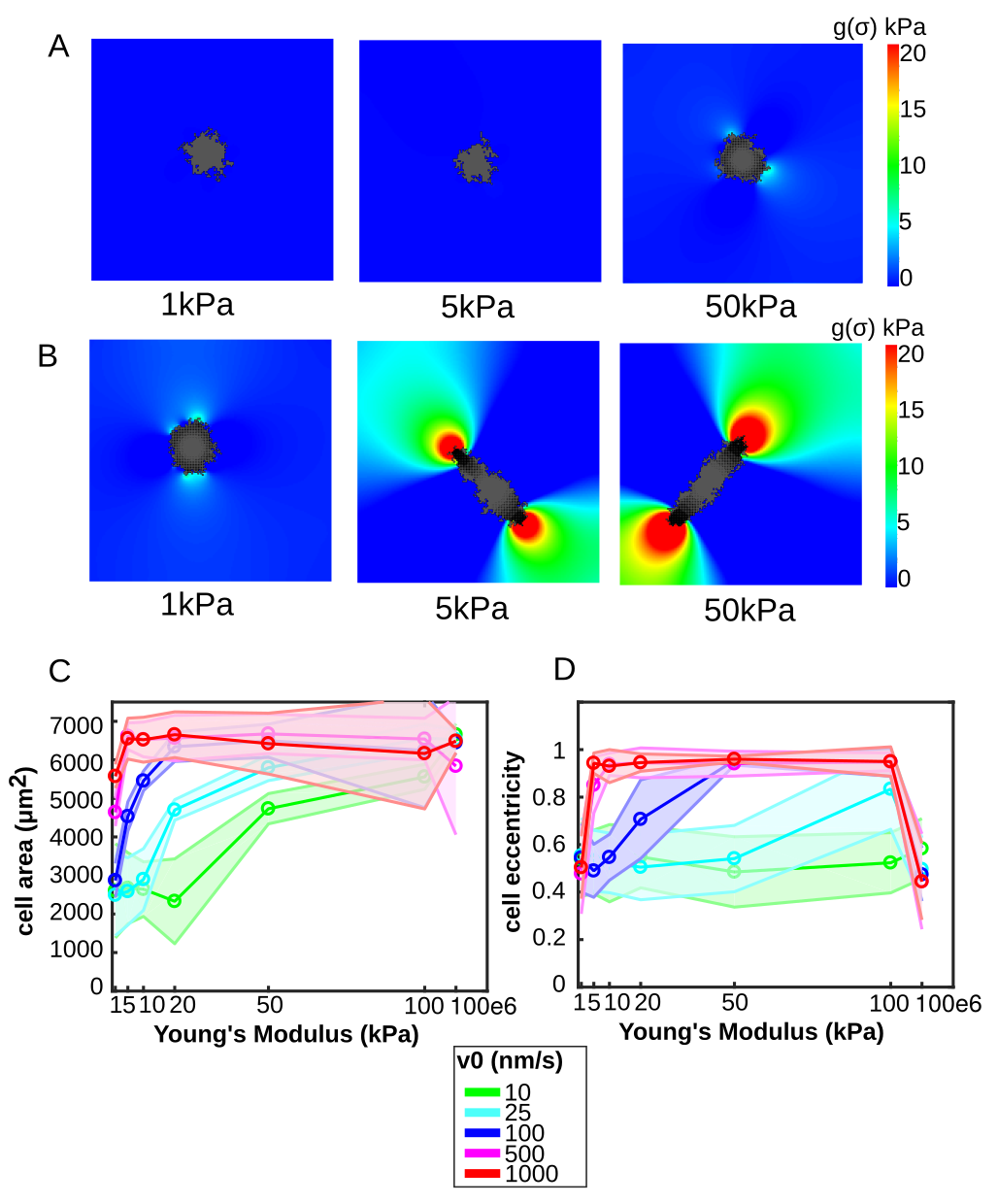}
\caption{Range of stiffness on which cells elongate depends on myosin motor velocity. Model M2 was used. (A) Example configurations of cells at 2000 MCS on substrates of 1,50 and 50 kPa with motor velocity 10 nm/s; (B) Example configurations of cells at 2000 MCS on substrates of 1,50 and 50 kPa with motor velocity 1000 nm/s. Colors (A-B): hydrostatic stress; (C) Mean cell area as a function of motor velocity, error bars: standard deviations over 25 simulations; (D) Mean cell eccentricity as a function of motor velocity, error bars: standard deviations over 25 simulations.}
\label{fig:fig4}
\end{figure}

Figure~\ref{fig:fig4}A and B shows the cell configurations for a slow (10 nm/s) and fast motor velocity (1000 nm/s), compared to the default value of 100 nm/s as shown in Figure~\ref{fig:fig3}A, respectively. This shows that cells with slow motors do not spread significantly and do not elongate, even on stiffer substrates. In contrast, cells with fast motors already spread more and elongate on softer matrices. We quantified this further by running 25 simulations for each combination of substrate stiffness and motor velocity. Figure~\ref{fig:fig4}B and Figure~\ref{fig:fig4}C plot the cell area and eccentricity, respectively, as a function of motor protein velocity. With the fastest velocity tested here (1000 nm/s), cell area saturates already at 5 kPa and cells elongate on a larger stiffness regime (5 kPa - 100 kPa). With the slowest motor velocity (10 nm/s), cells do not elongate at all, while they still spread well on stiff matrices. This is explained as follows. Decreasing $v_0$ is very similar to decreasing the stiffness of the substrate, because they both contribute to the rate of force build-up in the same way, given by $\frac{|\vec{F}_s|}{v_0 K}$. So, in terms of cell area, cells with slower motor proteins would obtain a larger spreading area at stiffer matrices. However, they are not able to elongate because forces are not built up fast enough to generate high enough matrix stress that induces the focal adhesion strengthening.

So, in summary, we predict that cells with faster motor proteins start spreading/elongating at softer substrates, while cells with slower motor proteins need a stiffer substrate to instigate a response.

\subsection*{Durotaxis explained by a bias in integrin clustering}

\begin{figure}
\centering
\includegraphics[width=\textwidth]{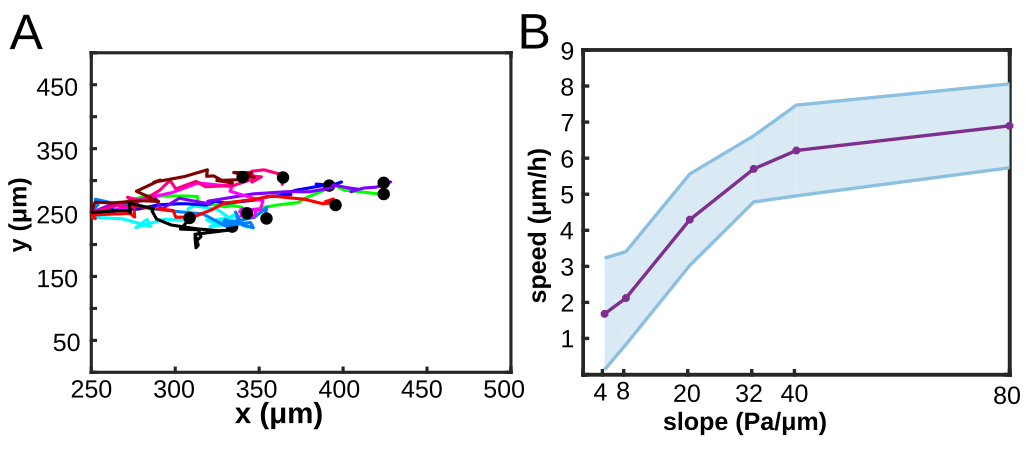}
\caption{Durotaxis as a result of integrin catch-bond dynamics. (A) Ten trajectories of durotacting cells on a matrix with slope 20 kPa/ $\mu$ m; (B) Cell speed as a function of the slope of the stiffness gradient.}
\label{fig:fig5}
\end{figure}

On substrates with a stiffness gradient, cells move up the stiffness gradient, a phenomena called durotaxis. Cells may durotact by sending out protrusions which better stick to stiff substrates because focal adhesions grow on stiff substrates \cite{Lo:2000cj,Wong2014}. Here, we investigate if force induced focal adhesion growth is sufficient to reproduce durotaxis. We simulated durotaxis by placing an initial circular cell with its center at $x$=$y$=250 $\upmu$m on a lattice of 1250 $\upmu$m by 500 $\upmu$m for 10000 MCS ($\approx 28h)$. In the $x$-direction, we let the stiffness increase from 1 kPa to 26 kPa, so with a slope of 20 Pa/$\upmu$m. Figure~\ref{fig:fig5}A plots ten different trajectories of the cell, showing that most cells have moved significantly in the $x$-direction, up the stiffness gradient. One realization is shown in Video S4. Cells, on average, move in the $x$-direction with a constant speed of around 4.3 $\upmu$m/h, measured as the slope of the $x$-coordinate of the cell from 25 simulations. Vincent \textit{et al.} \cite{Vincent2013} found speeds of 6.2 $\upmu$m/h with gradient slope 10 Pa/$\upmu$m \textit{in vitro} for mesenchymal stem cells. In our model, how far cells can move up the gradient, depends on the flexibility and motility of the cell. We varied $\lambda$, the Lagrangian multiplier of the area constraint, controlling cell flexibility, and the cellular temperature $T$, and found that both affect cell speed (Supplementary Figure~{8}). 

In the CPM, cell movement is a result of subsequent protrusions and retractions. In stiffer areas the focal adhesions grow larger, so that retraction are more likely to be made at more flexible parts of the matrix. As a result, the cell moves up the stiffness gradient.
So, naturally, one would expect that durotaxis depends on the slope of the stiffness gradient. Figure~\ref{fig:fig5}B shows the speed of the cell as a function of the slope of the stiffness gradient. Indeed, simulated cells move faster up the gradient if the slope is steeper, as observed in experimental conditions \cite{Vincent2013,Isenberg2009}. This is because the difference in focal adhesion growth between the front and the back of the cell is larger with a higher slope, causing a larger bias. We suspect that the durotaxis speed saturates at steep slopes, because the growth rate of focal adhesions is limited.

In conclusion, durotaxis is an emergent behavior in our model, cells exhibit durotaxis as a result of a biased growth of focal adhesions. A cell can build up forces faster on stiffer matrices, allowing focal adhesions to grow larger here. So, the cell better attaches at the stiffer part and will retract at the softer side. As a result, the cell moves up the stiffness gradient.


\subsection*{Discussion}
We have presented a multiscale computational model to show that force induced focal adhesion dynamics can explain 1) cell area increasing with substrate stiffness (Figure~\ref{fig:fig2}A-B), 2) cell elongation on substrates of intermediate stiffness (Figure~\ref{fig:fig3}A-B) and 3) durotaxis (Figure~\ref{fig:fig5}A). The model described cells spreading on an elastic substrate via focal adhesions, which are modelled as integrin clusters. Cells applied traction forces on integrin clusters, which grow according to catch-slip bond dynamics \cite{Novikova2013}. In this model, the disassembly of focal adhesions decreases with force. How fast a cell in our model can build up this force was assumed to depend on the stiffness of the matrix, based on a model by Schwarz et al. \cite{Schwarz2006}. On soft matrices, forces build up slowly such that integrin clusters do not have enough time to grow, while on stiff matrices forces build up fast such that integrin clusters can grow in size. Because we assumed that larger focal adhesions detach less likely from the substrate than smaller ones, cell spreading area increased on stiffer substrates (Figure~\ref{fig:fig2}B). If we included a feedback between matrix stresses and cell-matrix adhesive forces, simulated cells were able to elongate (Figure~\ref{fig:fig3}A-B). The model suggests that the range of substrate stiffness on which cells elongate depends on the velocity of the myosin molecular motors, which determine the rate of force build-up. Cells with higher motor protein velocity started to elongate on softer matrices (Figure~\ref{fig:fig4}). Finally, our model explains durotaxis as a bias in focal adhesion growth on stiffer matrices. Because extensions are more likely to stick at these regions and retractions are more likely to be made on the softer side, cells obtain a bias in cell motility up the stiffness gradient (Figure~\ref{fig:fig5}A). Our model predicted that cell velocity increases with the slope of the stiffness gradient (Figure~\ref{fig:fig5}B), which compares well with experimental data \cite{Isenberg2009,Vincent2013}. The spreading dynamics in our model also qualitatively compare well with \textit{in vitro} dynamics: the spreading dynamics in Figure~\ref{fig:fig2}C are similar to spreading area curves found \textit{in vitro} \cite{ReinhartKing:2005dq,Yeung2005} and the dynamics of cell elongation (Movie S1) resemble \textit{in vitro} observations \cite{Prager2011}.

\subsubsection*{Mechanisms driving cell elongation}
We hypothesized that the stabilization of focal adhesions by matrix stress is due to stretching of talin. Stretching of talin exposes vinculin binding sites \cite{Rio2009} and vinculin in turn binds the focal adhesion to the cytoskeleton, which strengthens cell-matrix adhesion \cite{Gallant2005}. Our model suggests that this might regulate cell elongation. In agreement with this observation, vinculin regulates cell elongation on glass substrates \cite{Ezzel1997}. We could attempt to further unravel how vinculin drives cell elongation by studying vinculin depleted cells on substrates of different stiffness, or by adapting talin in such a way that vinculin cannot bind as a result of talin stretching.

Interestingly, our model suggests that cells can also elongate if matrix stress induces an increase in cell traction forces (Supplementary Figure~{7}). This mechanism can be justified by two experimental observations; 1) vinculin increases cell traction forces \cite{Dumbauld2013} and 2) stressing focal adhesions induces $\alpha$-smooth muscle actin recruitment to stress fibers that in turn increases traction forces \cite{Pittet2006}. Experimental testing can be done to elucidate which mechanism might be required for cell elongation, since our model does not differentiate between these two and vinculin adhesion strengthening.

Our model also predicts that cells elongate on different ranges of substrate stiffness, due to different velocities of their myosin motors (Figure~\ref{fig:fig4}). This could explain why different cell types elongate on different stiffness regimes \cite{Yeung2005,Georges2005}, as they might express different isoforms of myosin motors. Many studies of different types of cells on compliant substrates have been performed, but often either the range of substrate stiffness tested differs or the type of matrix (\textit{i.e.} type of ligand, ligand density, or gel type) is different. Therefore, spreading of different cell types cannot be compared one to one. Model validation would benefit from more systematic \textit{in vitro} experiments of different cell types on compliant matrices. To then confirm this model prediction, it could be measured which isoform of myosin the cells express. There are some experiments that seem to support our model prediction. For instance, cell elongation is promoted in Dlc1 deficient ovarian tumour \cite{Sabbir2016}. Dlc1 leads to increases of phosphorylation level of nonmuscle IIA mysosin \cite{Sabbir2016}, which suggests that an increase in motor protein velocity indeed enables cells to elongate more. Furthermore, cells treated with blebbistatin on stiff matrices obtain phenotype as if they are on a soft matrix \cite{Jiang2016}, while upregulating myosin gives opposite results. In this paper by Jiang et al. \cite{Jiang2016} it was suggested that the actomyosin pulling speed produce has a similar effect on integrin stem cell lineage specification (which is highly associated with cell shape \cite{Mcbeath2004}) as the effective spring constant of the substrate. 

\subsubsection*{Focal adhesion regulation of cell spreading}
In our model, we have differentiated between integrin size dependent focal adhesion strength and focal adhesion strength reinforcement by structural proteins, which have been observed experimentally to be different mechanisms \cite{Roca-cusachs2009}. The strength of our model is that we can directly associate cell response to matrix stiffnes with mechanisms at the level of focal adhesions. Previous mathematical models were often based on how matrix stiffness influences mechanisms at the cellular level. The assumptions on how matrix stiffness affect cellular dynamics were however often motivated by a change in focal adhesion dynamics (stiffness sensing: \cite{Bischofs2003,Ni2007,Zemel2010,Kabaso2011}, durotaxis: \cite{Dokukina2010,Stefanoni2012,Allena2016,Yu2017}.) Similarly, the mechanism for cell spreading proposed in our previous work \cite{Oers:2014pcb}, was based on focal adhesion dynamics and depends on cell traction and cell adhesion \cite{Ramos2018}. In this previous model, we suggested that protrusions are more likely to stick and protrude to highly strained matrices that have strain-stiffened. This was motivated by the experimental observation that cells more efficiently build up forces on stiff matrices, which enables stabilization of focal adhesions \cite{Pelham:1997uc}. 

Cell based models including mechanosensitive focal adhesions have been used to study cell behavior (for instance, stepping locomotion \cite{Copos2017} and cell migration under the influence of external cues \cite{Uatay2018}). In terms of cell spreading, hybrid cell-focal adhesion models suggested that cell spreading requires focal adhesions to upregulate cell forces \cite{Ronan2014,Vernerey2014}. In our model, such a feedback is only required for cells to elongate but not for an increase in cell spreading area. In line with this view, the model by Stolarska et al. \cite{Stolarska2017} suggested that the mechanosensitive growth of focal adhesions alone could not explain increased cell spreading on stiff matrices \cite{Stolarska2017} because increased cell contraction on stiff matrices resists cell spreading. In contrast to these previous models, our model suggests that a more rapid build up of forces on stiff matrices, allows focal adhesions to stabilize enough to enable cell spreading and no increased traction force is required. Cells in our model are able to spread on rigid matrices without focal adhesion reinforcement of cell traction forces, because the adhesion strength of large focal adhesions resist cell retractions on stiff matrices. 

\subsubsection*{Focal adhesion regulation of durotaxis}
We also compare our results to recent hybrid cell-focal adhesion models that were used to study durotaxis. In the model by Yu et al.  \cite{Yu2017}, the number of focal adhesions was assumed to be higher on stiff substrates and the distribution of focal adhesions was assumed to be more narrow on stiff substrates. Both the number and distribution of focal adhesion then controlled the deviation from the direction of motion: on stiff matrices cells move more persistent, causing it to durotact. Recently, it was also proposed that cells durotact by tugging on the matrix and then polarizing towards areas that the cell perceives as stiff \cite{Kim2018}. So, both these models suggest that the mechanosensitivity of focal adhesions drive durotaxis by polarizing the cell. Feng et al. \cite{Feng2018} showed that if focal adhesion degradation is higher in the back than in the front \textit{and} focal adhesions mature under applied force, then a cell can durotact. Based on experimental observations, Novikova and et al., presumed that cells move more persistently on stiffer substrates and showed that a persistent random walk can reproduce durotaxis \cite{Novikova2017}.  In contrast to previous models, our model suggests that durotaxis emerges from the mechanosensitive growth of focal adhesions and that no inherent polarized or persistent cell migration is required. 

\subsubsection*{Model limitations}
A limitation of our model is that it cannot accurately predict increasing focal adhesion sizes as a function of substrate stiffness (see Figure~\ref{fig:fig2}D), while this has been observed experimentally \cite{Prager2011}. This may be explained by modeling choices. In the CPM, cells only make retractions at the boundary of the cell, so in the middle of the cell, integrin clusters continue to grow even on soft matrices. Also, there is a fixed pool of free integrin bonds, making the growth rate of new focal adhesions to go down due to existing focal adhesions. Furthermore, our lattice based model does not define spatial effects in integrin clustering. In reality, small clusters may merge into larger adhesions and the availability of integrins that can bind to ECM, active integrin, is spatially and temporally regulated. Cells produce integrins, that diffuse and are activated within the cell. This activation of integrin depends on interaction with other proteins, such as talin \cite{Welf2012} and vinculin \cite{Humphries2007}. Furthermore, Stretching of p130cas induces its phosphorylation, which in turn activates the small GTPase Rap1 \cite{Sawada2006} which activates integrins \cite{Bos2003}. So, to better reproduce focal adhesion growth in future models, we can include other relevant mechanisms such as diffusion and the activation of integrins \cite{DESHPANDE20081484,Block2011,Welf2012,Vernerey2014}. However, because we were interested in cell shape in this work, which can be predicted with our model, we find the level of detail of focal adhesion dynamics sufficient at the moment.

\subsubsection*{Conclusion}
In summary, we propose that the mechanosensitive response of molecules in focal adhesions suffice to explain the response of cells to matrix stiffness. In agreement with experimental observations, cells spread more on stiff matrices and obtain an elongated shape if the matrix is stiff enough. Furthermore, cells durotact and move faster with steeper stiffness gradients. This model paves the way to study how specific molecular mechanisms within focal adhesions impact cell and tissue level responses to matrix mechanics. This can give rise to new targets of treatment and the design of tissue engineering experiments.


\section*{Methods}

We developed a multiscale model where cell movement depends on force induced focal adhesion dynamics. The model couples a cell-based model, substrate model and focal adhesion model in the following way. The Cellular Potts Model (CPM) describes cell movement. The shape of the cell is used to describe the stall forces that the cell exerts on the focal adhesions attached to a flexible substrate. These forces affect the growth of the focal adhesions. We assume that focal adhesions are clusters of integrins that behave as catch-slip bonds. Its dynamics are described using ordinary differential equations (ODEs). Finally, we assume that the cell-matrix link is strengthened by matrix stresses, which we calculate using a finite element model (FEM). The default parameter values are described in Supplementary Table~{S1}. 

\subsection*{Cellular Potts Model}
To simulate cell movement, we used the Cellular Potts Model (CPM) \cite{Graner:1992ve}. The CPM describes cells on a lattice $\Lambda \subset \mathbb{Z}^2$ as a set of connected lattice sites. Since the simulations in this article are limited to one cell, we describe the CPM here for a single cell. To each lattice site $\vec{x} \in \Lambda$ a spin $s(\vec{x}) \in \lbrace 0,1 \rbrace$ is assigned. This spin value indicates if $\vec{x}$ is covered by the cell, $s(\vec{x})=1$, or not, $s(\vec{x})=0$. Thus the cell is given by the set,
\begin{equation}C = \lbrace \vec{x}\in\Lambda | s(\vec{x})=1 \rbrace.
\label{eq:cellset}
\end{equation}
The cell set $C$ evolves by dynamic Monte Carlo simulation. During one Monte Carlo Step (MCS), the algorithm attempts copy a spin value $s(\vec{x})$ from a source site $\vec{x}$ into a neighboring target site $\vec{x}\prime$ from the usual Moore neighbourhood. Such copies mimic cellular protrusions and retractions. During an MCS, $N$ copy attempts are made, with $N$ the number of lattice sites in the lattice. Whether a copy is accepted or not depends on a balance of forces, which are represented in a Hamiltonian $H$. 

A copy is accepted if $H$ decreases, or with a Boltzmann probability otherwise, to allow for stochasticity of cell movements:
\begin{equation}
P(\Delta H)= 
\begin{cases}
1 & \textrm{if}\;\Delta H + Y <0 \\
e^{-(\Delta H+ Y)/T} & \textrm{if}\;\Delta H + Y\geq 0.
\end{cases}
\label{eq:boltzmann}
\end{equation}
Here ${\Delta H} = H_{\mathrm{after}}-H_{\mathrm{before}}$ is the change in $H$ due to copying, and the cellular temperature $T \geq 0$ determines the extent of random cell motility. Furthermore, $Y$ denotes a yield energy, an energy a cell needs to overcome to make a movement. Finally, to prevent cells from splitting up into disconnected patches, we use a connectivity constraint that always rejects a copy if it would break apart a cell in two or more pieces.

Following Ref.~\cite{Albert2014}, we use the following Hamiltonian:
\begin{equation}
H =  \underbrace{\lambda A^2}_{\text{contraction}} + \underbrace{\sum_{\text{neighbours} (\vec{x},\vec{x}\prime) } {J(s(\vec{x}),s(\vec{x}\prime))}}_{\text{line tension}} - \underbrace{\lambda_C \frac{A}{A_h + A}}_{\text{cell-matrix adhesion}}.
\end{equation}
The first term of $H$ denotes cell contraction, where $A$ is the area of the cell and $\lambda$ is the corresponding Lagrange multiplier. In the second term, $J(s(\vec{x}),s(\vec{x}\prime)$ are the adhesive energy between two sites $\vec{x}$ and $\vec{x}\prime$ with spins $s(\vec{x})$ and $s(\vec{x}\prime)$. When taking a sufficient large neighborhood, the second term describes a line tension, as it approximates the perimeter of a cell \cite{Magno2015}. We take a neighborhood radius of 10 for this calculation. The third term describes the formation of adhesive contacts of cells with the substrate, where the bond energies lower the total energy \cite{Albert2014}, causing the cells to spread. The parameter $\lambda_\mathrm{C}$ is the corresponding Lagrange multiplier. The energy gain of occupying more lattice sites saturates with the cell area, as the total number of binding sites is limited. The parameter $A_h$ regulates this saturation. 

To describe cell-matrix binding via focal adhesions, we implement the following yield energy in the CPM
\begin{equation}
Y = \lambda_\mathrm{N} \frac{N(\vec{x}')-N_0}{N_h + N(\vec{x}')} \cdot \mathbf{1}_{s(\vec{x}')=1} \cdot \mathbf{1}_{s(\vec{x})=0},
\label{eq:yield}
\end{equation}
where $N(\vec{x}')$ is the size of the focal adhesion at the target site. This models that a retraction is energetically costly for a cell to make, because it needs to break the actin-integrin connection. We assume that the size of the actin-integrin link is proportional to the size of the focal adhesion, \textit{i.e.} the number of integrin bonds \cite{Boettiger2007}, and that the strength of a focal adhesion saturates \cite{Gallant2005} with a parameter $N_h$. The substraction of $N_0$ represents that a focal adhesion only creates extra linkage if it is greater than a nascent adhesion. Note that the $Y$ can not become negative, because we assume that focal adhesions smaller than $N_0$, a nascent adhesion, breaks down due to its short lifetime, see section `Focal Adhesions'. So, only focal adhesions larger than $N_0$ create a yield energy. In section `Substrate stresses', we further adapt this yield energy to describe a matrix stress induced focal adhesion reinforcement.

\subsection*{Cell traction forces}
Following Schwarz et al. \cite{Schwarz2006}, we assume that traction forces are generated by myosin molecular motors on the actin fibers, of which the velocity is given by
\begin{equation}
v(\vec{F}) = v_0 \left ( 1 - \vec{F}/\vec{F_s} \right ),
\end{equation}
where $v_0$ is a free velocity. The traction forces are applied to the ECM, which we assume is in plane stress. The constitutive equation is given by $h \vec{\nabla} \dund{\sigma} =\vec{F}$ where $\dund{\sigma}$ is the ECM stress tensor and $h$ is the thickness of the ECM. We assume that the ECM is isotropic, uniform, linearly elastic and we assume infinitesimal strain theory. We solve this equation using a Finite Element Model (FEM) (see section `Substrate stresses'). In the FEM, traction field $\vec{f}$ and ECM deformation $\vec{u}$ are related by:
\begin{equation}
\mathbf{K} \vec{u} = \vec{f},
\end{equation}
where $\mathbf{K}$ is the global stiffness matrix given by assemblying the local stiffness matrices $\mathbf{K}_e$ for each lattice site $e$
\begin{equation}
\mathbf{K}_e= h \int \mathbf{B}^T \frac{E}{1-\nu^2} \begin{pmatrix} 1 & \nu & 0 \\ \nu & 1 & 0 \\ 0 & 0 & \frac{1-\nu}{2}   \end {pmatrix} \mathbf{B},
\end{equation}
where $\mathbf{B}$ is the conventional strain-displacement matrix for a four-noded quadrilateral element \cite{Davies:2011wr} and $E$ is the Young's modulus and $\nu$ is the Poisson's ratio of the ECM. For more details on this part of the model, we refer to our previous work \cite{Oers:2014pcb,Rens2017}. 

Following Schwarz \textit{et al.} \cite{Schwarz2006}, the force build-up is given by the ODE:
\begin{equation}
\mathbf{K} \vec{v}(\vec{f}) = \frac{d\vec{f}}{dt},
\end{equation}
The matrix $\mathbf{K}$ describes force interactions between neighbouring nodes in the ECM. However, since solving this equation is expensive, we ignore the interactions between neighbouring sites, \textit{i.e.}, we reduce $\mathbf{K}$ to  a scalar for each site $\vec{x}$. This gives us,
\begin{equation}
\vec{F}(\vec{x},t) = \vec{F}_s(\vec{x}) + (\vec{F}_0(\vec{x})-\vec{F}_s(\vec{x}))\exp({-t/t_k}),
\label{eq:forcebuildup}
\end{equation}
where $\vec{F}_0$ is the force already exerted by the actin and $t_k=\frac{|\vec{F}_s|}{v_0 K}$. Here, $K$ is given by the diagonal entry of $\mathbf{K}$ at site $\vec{x}$, i.e. the stiffness of this node, neglecting changes in local stiffness due to connections to neighbouring nodes described in the off-diagonal entries of $\mathbf{K}$. 
Since the cell configuration and therefore the traction forces change each MCS, the tension on the focal adhesions does not build up from zero, but from the tension that was built up during the previous MCS: $\vec{F}_0$ at the current MCS is given by $\vec{F}(t_\mathrm{FA})$ of the previous MCS. 

To calculate the stall force of the actin fibers, $\vec{F}_s$, we employ the empirical first-moment-of-area (FMA) model \cite{Lemmon:2010ju}. This model infers the stall forces from the shape of the cell of the CPM, based on the assumption that a network of actin fibers in the cell acts as a single, cohesive unit, 
\begin{equation}
\vec{F}_s(\vec{x})=\frac{\mu}{A} \sum_{\{\vec{y}\in C | [\vec{y}\vec{x}] \subset C \}} \vec{x}-\vec{y}.
\end{equation}
So, the force at site $\vec{x}$ is calculated as the sum of forces between $\vec{x}$ and all other sites $\vec{y}$ within cell $C$ that are connected to $\vec{x}$ (this sum excludes line segments $[\vec{y}\vec{x}]$ running outside the cell that occur if the shape of cell $C$ is non-convex). The force is assumed to be proportional to the distance between the sites. 
We divide over the cell area $A$ such that force increases roughly linear with cell area, as experimentally observed \cite{Califano:2010dp}. 

\subsection*{Focal adhesions}
At each lattice site occupied by the cell, $\vec{x}\in C$,  a focal adhesion is modeled as a cluster of bound integrin bonds $N$. Each individual integrin bond behaves as a catch-slip bond, whose lifetime is maximal under a positive force \cite{Novikova2013}. Accordingly, the growth of a cluster of such bonds is described by the ODE-model derived by Novikova and Storm \cite{Novikova2013}, 
\begin{equation}
\frac{d N(\vec{x},t)}{d t} = \gamma N_a(t)\left(1-\frac{N(\vec{x},t)}{N_b}\right) - d(\phi(\vec{x},t))N(\vec{x},t) \label{eq:ode}
\end{equation}
with $\gamma$ is the binding rate of integrins to the ECM, $N_a$ the number of free bonds, and $N_b$ the maximum number of bound bonds a lattice site can contain. This logistic growth term is a slight adaptation compared to Novikova and Storm \cite{Novikova2013}. This additional term was added to avoid packing more integrins in a lattice site, than the size of a lattice site ($\Delta x^2$) can accomodate for. The degradation of the focal adhesions $d(\phi)$ depends on the tension $\phi$ on the focal adhesion $N$. This degradation rate is given by
\begin{equation}
d(\phi(\vec{x},t)) = \exp{\left(\frac{\phi(\vec{x},t)}{N(\vec{x},t)}-\phi_s\right)} + \exp{\left(-\frac{\phi(\vec{x},t)}{N(\vec{x},t)}+\phi_c\right)}
\label{eq:dstress}
\end{equation}
 where $\phi_s$ and $\phi_c$ describe the slip and catch bond regime in N/m$^2$, respectively. Here, $\phi(\vec{x},t)=\frac{|\vec{F}(\vec{x},t)|}{\Delta x^2}$ is the stress applied to the lattice site of the focal adhesion. We assume that the number of free bonds $N_a$ is limited by the number of available integrin receptors in the entire cell, $N_m$. These $N_m$ receptors can be recruited to each focal adhesion site and enable binding of a bond. Thus, $N_a(t) = N_m - \sum_{\vec{x}\in C} N(\vec{x},t)$. We let the focal adhesions grow after each MCS for $t_\mathrm{FA}$ seconds with time increments of $\Delta t_\mathrm{FA}$. If there is no pre-existing focal adhesion at site $\vec{x} \in C$, we set $N(\vec{x})=N_0$, so that at this site, a new initial adhesion is formed. This assumption represents the generation of focal complexes or nascent adhesions, precursors of focal adhesions that contain a small amount of integrins and have a very short lifetime\cite{Changede2016}. Also, after the focal adhesions were allowed to grow, \textit{i.e.} after $t=t_\mathrm{FA}$ seconds, we set all $N(\vec{x})<N_0$ back to $N(\vec{x})=N_0$, again modeling the quick (re)generation of nascent adhesions.

When a site $\vec{x}$ is removed from the cell $C$ after a retraction, such that $s(\vec{x})=0$,  we set $N(\vec{x})=0$ reflecting the destruction of the focal adhesion. We assume that if a cell extends, i.e. a site $\vec{x}$ is added to the cell $C$, a nascent adhesion is formed: we set $N(\vec{x})=N_0$. 

\subsection*{Substrate stresses}
The forces that were build up during a MCS, $\vec{F}(t_\mathrm{FA})$ are applied as planar forces to a finite element model (FEM). The FEM calculates the stress tensor $\dund{\sigma}(\vec{x})$ on each lattice site. We assume that the integrin-cytoskeletal adhesion strengthens as a result of stress. We define
\begin{equation}
g(\dund{\sigma})=
\begin{cases}
\frac{1}{2}(\sigma_{xx}+\sigma_{yy}) & \text{if } \frac{1}{2}(\sigma_{xx}+\sigma_{yy}) \geq 0\\
0 & \text{if } \frac{1}{2}(\sigma_{xx}+\sigma_{yy})<0
\end{cases}
\end{equation}
the positive hydrostatic stress of the stress tensor that describes how much stress the focal adhesion experiences. Now, we extend the yield energy as follows:
\begin{equation}
Y = \lambda_\mathrm{N} \frac{N(\vec{x}')-N_0}{N_h + N(\vec{x}')} \cdot \left ( 1+p \frac{g(\dund{\sigma}(\vec{x}'))}{\sigma_h+g(\dund{\sigma}(\vec{x}'))} \right ) \cdot \mathbf{1}_{\vec{x}\prime\in C} \cdot \mathbf{1}_{\vec{x}\notin C} \label{eq:yield2}
\end{equation}
We thus assume that stress strengthens the focal adhesions, with parameter $p$ and that this strengthening saturates with parameter $\sigma_h$. 

\subsection*{Stiffness gradient}
To study durotaxis, we model a stiffness gradient in the $x$-direction on a lattice of 1250 $\upmu$m by 500 $\upmu$m. The Young's modulus of the substrate $E (Pa)$ is given by $E(x)=\mathrm{max} \lbrace 1 ,6000+(x-250) \cdot \mathrm{slope} \rbrace$, with $x$ in $\upmu m$, such that the Young's modulus at the center of the cell at time $t=0$ is 6000 Pa and is nonzero. The default value for the slope is 20 Pa/$\upmu$m.

\section{Acknowledgments}
This work was part of the research program Innovational Research Incentives Scheme Vidi Cross-divisional 2010 ALW with project number 864.10.009 to RMHM, which is (partly) financed by the Netherlands Organization for Scientific Research (NWO).

\bibliographystyle{pnas-new.bst}
\bibliography{FA_draft.bib}

\begin{thebibliography}{10}

\bibitem{Califano2010}
Califano JP, Reinhart-King CA (2010) Substrate stiffness and cell area predict
  cellular traction stresses in single cells and cells in contact.
\newblock {\em Cellular and Molecular Bioengineering} 3(1):68--75.

\bibitem{Pelham:1997uc}
Pelham RJR, Wang YlY (1997) Cell locomotion and focal adhesions are regulated
  by substrate flexibility.
\newblock {\em Proceedings of the National academy of Sciences of the United
  States of America} 94(25):13661--13665.

\bibitem{Ghibaudo2008}
Ghibaudo M, et~al. (2008) Traction forces and rigidity sensing regulate cell
  functions.
\newblock {\em Soft Matter} 4(9):1836--1843.

\bibitem{Engler2004}
Engler A, et~al. (2004) Substrate compliance versus ligand density in cell on
  gel responses.
\newblock {\em Biophysical Journal} 86(1):617--628.

\bibitem{Mullen2015}
Mullen CA, Vaughan TJ, Billiar KL, Mcnamara LM (2015) The effect of substrate
  stiffness, thickness, and cross-linking density on osteogenic cell behavior.
\newblock {\em Biophysical Journal} 108(7):1604--1612.

\bibitem{Lo:2000cj}
Lo CM, Wang HB, Dembo M, Wang YL (2000) Cell movement is guided by the rigidity
  of the substrate.
\newblock {\em Biophysical Journal} 79(1):144--152.

\bibitem{Isenberg2009}
Isenberg BC, Dimilla PA, Walker M, Kim S, Wong JY (2009) Vascular smooth muscle
  cell durotaxis depends on substrate stiffness gradient strength.
\newblock {\em Biophysical Journal} 97(5):1313--1322.

\bibitem{Vincent2013}
Vincent L, Choi Y, Alonso-Latorre B, del Alamo J, Engler A (2013) Mesenchymal
  stem cell durotaxis depends on substrate stiffness gradient strength.
\newblock {\em Biotechnology Journal} 8(4):472--484.

\bibitem{Jansen2015}
Jansen KA, et~al. (2015) A guide to mechanobiology: Where biology and physics
  meet.
\newblock {\em Biochimica et Biophysica Acta} 1853(11 Pt B):3043--3052.

\bibitem{Bershadsky2003}
Bershadsky AD, Balaban NQ, Geiger B (2003) Adhesion-dependent cell
  mechanosensitivity.
\newblock {\em Annual Review Cell Developmental Biology} 19:677--695.

\bibitem{ZaidelBar:2007hr}
Zaidel-Bar R, Itzkovitz S, Ma'ayan A, Iyengar R, Geiger B (2007) {Functional
  atlas of the integrin adhesome.}
\newblock {\em Nature Cell Biology} 9(8):858--867.

\bibitem{WinogradKatz:2014dg}
Winograd-Katz SE, F{\"a}ssler R, Geiger B, Legate KR (2014) {The integrin
  adhesome: from genes and proteins to human disease}.
\newblock {\em Nat Rev Mol Cell Biol} 15(4):273--288.

\bibitem{Ezzel1997}
Ezzel R, Goldmann W, Wang N, Parashurama N, Ingber D (1997) Vinculin promotes
  cell spreading by mechanically coupling integrins to the cytoskeleton.
\newblock {\em Experimental Cell Research} 231(1):14--26.

\bibitem{Prager2011}
Prager-Khoutorsky M, et~al. (2011) Fibroblast polarization is a
  matrix-rigidity-dependent process controlled by focal adhesion
  mechanosensing.
\newblock {\em Nature Cell Biology} 13(12):1457--1465.

\bibitem{Broussard2008}
Broussard JA, Webb DJ, Kaverina I (2008) Asymmetric focal adhesion disassembly
  in motile cells.
\newblock {\em Current Opinion in Cell Biology} 20(1):85--90.

\bibitem{Plotnikov:2012cp}
Plotnikov SV, Pasapera AM, Sabass B, Waterman CM (2012) Force fluctuations
  within focal adhesions mediate ecm-rigidity sensing to guide directed cell
  migration.
\newblock {\em Cell} 151(7):1513--1527.

\bibitem{Chen2013}
Chen Y, Pasapera AAM, Koretsky AP, Waterman CM (2013) Orientation-specific
  responses to sustained uniaxial stretching in focal adhesion growth and
  turnover.
\newblock {\em Proceedings of the National academy of Sciences of the United
  States of America} 110(26):E2352--61.

\bibitem{Kim2013}
Kim HY, Varner VD, Nelson CM (2013) Apical constriction initiates new bud
  formation during monopodial branching of the embryonic chicken lung.
\newblock {\em Development} 140(15):3146--3155.

\bibitem{Rio2009}
Rio A, et~al. (2009) Stretching single talin rod molecules activates vinculin
  binding.
\newblock {\em Science} 323(5914):638--642.

\bibitem{Gallant2005}
Gallant ND, Michael KE, Garcı J (2005) Cell adhesion strengthening:
  Contributions of adhesive area, integrin binding, and focal adhesion
  assembly.
\newblock {\em Molecular Biology of the Cell} 16(9):4329--4340.

\bibitem{Kong2009}
Kong F, Mould AP, Humphries MJ, Zhu C (2009) Demonstration of catch bonds
  between an integrin and its ligand.
\newblock {\em The Journal of Cell Biology} 185(7):1275--1284.

\bibitem{Dembo1988}
Dembo M, Torney DC, Saxman K, Hammer D (1988) The reaction-limited kinetics of
  membrane-to-surface adhesion and detachment.
\newblock {\em Proceedings of the Royal Society of London. Series B, Biological
  sciences} 234(1274):55--83.

\bibitem{Bissell:1982wb}
Bissell MJ, Hall HG, Parry G (1982) {How does the extracellular matrix direct
  gene expression?}
\newblock {\em Journal of Theoretical Biology} 99(1):31--68.

\bibitem{Ni2007}
Ni Y, Chiang MYM (2007) Cell morphology and migration linked to substrate
  rigidity.
\newblock {\em Soft Matter} 3(10):1285--1292.

\bibitem{Shenoy2015}
Shenoy VB, Wang H, Wang X (2015) A chemo-mechanical free-energy-based approach
  to model durotaxis and extracellular stiffness-dependent contraction and
  polarization of cells.
\newblock {\em Interface Focus} 6(1):20150067.

\bibitem{Oers:2014pcb}
van Oers RFM, Rens EG, LaValley DJ, Reinhart-King CA, Merks RMH (2014)
  Mechanical cell-matrix feedback explains pairwise and collective endothelial
  cell behavior in vitro.
\newblock {\em PLoS Computational Biology} 10(8):e1003774.

\bibitem{Dokukina2010}
Dokukina IV, Gracheva ME (2010) A model of fibroblast motility on substrates
  with different rigidities.
\newblock {\em Biophysical Journal} 98(12):2794--2803.

\bibitem{Lazopoulos}
Lazopoulos~KA SD (2008) Durotaxis as an elastic stability phenomenon.
\newblock {\em Journal of Biomechanics} 41(6):1289--1294.

\bibitem{Harland2011}
Harland B, Walcott S, Sun SX (2011) Adhesion dynamics and durotaxis in
  migrating cells.
\newblock {\em Physical Biology} 8(1):015011.

\bibitem{Stefanoni2012}
Stefanoni F, Ventre M, Mollica F, Netti PA (2011) A numerical model for
  durotaxis.
\newblock {\em Journal of Theoretical Biology} 280(1):150--158.

\bibitem{Aubry2015}
Aubry D, Gupta M, Ladoux B, Allena R (2015) Mechanical link between durotaxis,
  cell polarity and anisotropy during cell migration.
\newblock {\em Physical Biology} 12(2):026008.

\bibitem{Allena2016}
Allena R, Scianna M, Preziosi L (2016) Mathematical biosciences a cellular
  potts model of single cell migration in presence of durotaxis.
\newblock {\em Mathematical Biosciences} 275:57--70.

\bibitem{Novikova2017}
Novikova EA, Raab M, Discher DE, Storm C (2017) Persistence-driven durotaxis:
  Generic, directed motility in rigidity gradients.
\newblock {\em Physical Review Letters} 118(7):078103.

\bibitem{Yu2017}
Yu G, Feng J, Man H, Levine H (2017) Phenomenological modeling of durotaxis.
\newblock {\em Physical Review E} 96:010402.

\bibitem{Kim2018}
Kim MC, Silberberg YR, Abeyaratne R, Kamm RD, Asada HH (2018) Computational
  modeling of three-dimensional ecm-rigidity sensing to guide directed cell
  migration.
\newblock {\em Proceedings of the National academy of Sciences of the United
  States of America} 115(3):E390--E999.

\bibitem{DESHPANDE20081484}
Deshpande VS, Mrksich M, McMeeking RM, Evans AG (2008) A bio-mechanical model
  for coupling cell contractility with focal adhesion formation.
\newblock {\em Journal of the Mechanics and Physics of Solids} 56(4):1484 --
  1510.

\bibitem{Ronan2014}
Ronan W, Deshpande VS, Mcmeeking RM, Mcgarry JP (2014) Cellular contractility
  and substrate elasticity: A numerical investigation of the actin cytoskeleton
  and cell adhesion.
\newblock {\em Biomechanics and Modeling in Mechanobiology} 13(2):417--435.

\bibitem{Vernerey2014}
Vernerey FJ, Farsad M (2014) A mathematical model of the coupled mechanisms of
  cell adhesion , contraction and spreading.
\newblock {\em Journal of Mathematical Biology} 68(4):989--1022.

\bibitem{Besser2006}
Besser A, Safran SA (2006) Force-induced adsorption and anisotropic growth of
  focal adhesions.
\newblock {\em Biophysical Journal} 90(10):3469--3484.

\bibitem{Stolarska2017}
Stolarska MA, Rammohan AR (2017) Center or periphery? modeling the effects of
  focal adhesion placement during cell spreading.
\newblock {\em PLOS ONE} 12(2):1--22.

\bibitem{Novikova2013}
Novikova EA, Storm C (2013) Contractile fibers and catch-bond clusters: A
  biological force sensor?
\newblock {\em Biophysical Journal} 105(6):1336--1345.

\bibitem{Schwarz2006}
Schwarz US, Erdmann T, Bischofs IB (2006) Focal adhesions as mechanosensors:
  The two-spring model.
\newblock {\em Biosystems} 83(2-3):225--232.

\bibitem{Rens2017}
Rens EG, Merks RMH (2017) Cell contractility facilitates alignment of cells and
  tissues to static uniaxial stretch.
\newblock {\em Biophysical Journal} 112(4):755--766.

\bibitem{Lemmon:2010ju}
Lemmon CA, Romer LH (2010) A predictive model of cell traction forces based on
  cell geometry.
\newblock {\em Biophysical Journal} 99(9):L78--L80.

\bibitem{Balcioglu1316}
Balcioglu HE, van Hoorn H, Donato DM, Schmidt T, Danen EHJ (2015) The integrin
  expression profile modulates orientation and dynamics of force transmission
  at cell{\textendash}matrix adhesions.
\newblock {\em Journal of Cell Science} 128(7):1316--1326.

\bibitem{Asanoe13281}
Asano S, et~al. (2017) Matrix stiffness regulates migration of human lung
  fibroblasts.
\newblock {\em Physiological Reports} 5(9):e13281.

\bibitem{ReinhartKing:2005dq}
Reinhart-King CA, Dembo M, Hammer DA (2005) The dynamics and mechanics of
  endothelial cell spreading.
\newblock {\em Biophysical Journal} 89(1):676--689.

\bibitem{Yeung2005}
Yeung T, et~al. (2005) Effects of substrate stiffness on cell morphology,
  cytoskeletal structure, and adhesion.
\newblock {\em Cell Motility and the Cytoskeleton} 60(1):24--34.

\bibitem{Paszek:2005eh}
Paszek MJ, et~al. (2005) Tensional homeostasis and the malignant phenotype.
\newblock {\em Cancer Cell} 8(3):241--254.

\bibitem{Califano:2010dp}
Califano JP, Reinhart-King CA (2010) Substrate stiffness and cell area predict
  cellular traction stresses in single cells and cells in contact.
\newblock {\em Cellular and Molecular Bioengineering} 3(1):68--75.

\bibitem{Dumbauld2013}
Dumbauld DW, et~al. (2013) How vinculin regulates force transmission.
\newblock {\em Proceedings of the National academy of Sciences of the United
  States of America} 110(24):9788--9793.

\bibitem{Pittet2006}
Goffin JM, et~al. (2006) Focal adhesion size controls tension-dependent
  recruitment of $\alpha$ -smooth muscle actin to stress fibers.
\newblock {\em The Journal of Cell Biology} 172(2):259--268.

\bibitem{Uyeda2011}
Uyeda TQP, Iwadate Y, Umeki N, Nagasaki A, Yumura S (2011) Stretching actin
  filaments within cells enhances their affinity for the myosin ii motor
  domain.
\newblock {\em PLOS ONE} 6(10):e26200.

\bibitem{Ma2012}
Ma X, Adelstein RS (2012) In vivo studies on nonmuscle myosin ii expression and
  function in heart development.
\newblock {\em Frontiers in Bioscience} 17:545--555.

\bibitem{Murakimi1993}
Sugiura T, Miyata H, Kawai Y, Matoba H, Murakami N (1993) Changes in myosin
  heavy chain isoform expression of overloaded rat skeletal muscles.
\newblock {\em The International Journal of Biochemistry} 25(11):1609--1613.

\bibitem{Norstrom2010}
Norstrom M, Smithback P, Rock R (2010) Unconventional processive mechanics of
  non-muscle myosin ii-b.
\newblock {\em Journal of Biological Chemistry} 285(34):26326--26334.

\bibitem{Kelley1996}
Kelley CA, et~al. (1996) Xenopus nonmuscle myosin heavy chain isoforms have
  different subcellular localizations and enzymatic activities.
\newblock {\em Journal of Cell Biology} 134(3):675--687.

\bibitem{Vogel2013}
Vogel SK, Petrasek Z, Heinemann F, Schwille P (2013) Myosin motors fragment and
  compact membrane-bound actin filaments.
\newblock {\em eLife} 2:e00116.

\bibitem{Wong2014}
Wong S, Guo Wh, Wang Yl (2014) Fibroblasts probe substrate rigidity with
  filopodia extensions before occupying an area.
\newblock {\em Proceedings of the National academy of Sciences of the United
  States of America} 111(48):1--6.

\bibitem{Georges2005}
Georges PC, Janmey PA (2005) Cell type-specific response to growth on soft
  materials.
\newblock {\em Journal of Applied Physiology: Respiratory, Environmental and
  Exercise Physiology} 98(4):1547--1553.

\bibitem{Sabbir2016}
Sabbir MG, Dillon R, Mowat MRA (2016) Dlc1 interaction with non-muscle myosin
  heavy chain ii-a ( myh9 ) and rac1 activation.
\newblock {\em Biology Open} 5(4):452--460.

\bibitem{Jiang2016}
Jiang L, et~al. (2016) Cells sensing mechanical cues: Stiffness influences the
  lifetime of cell-extracellular matrix interactions by affecting the loading
  rate.
\newblock {\em ACS Nano} 10(1):207--217.

\bibitem{Mcbeath2004}
Mcbeath R, Pirone DM, Nelson CM, Bhadriraju K, Chen CS (2004) Cell shape,
  cytoskeletal tension, and rhoa regulate stem cell lineage commitment.
\newblock {\em Developmental Cell} 6(4):483--495.

\bibitem{Roca-cusachs2009}
Roca-cusachs P, Gauthier NC, Sheetz MP (2009) Clustering of $\alpha$5$\beta$1
  integrins determines adhesion strength whereas $\alpha$v$\beta$3 and talin
  enable mechanotransduction.
\newblock {\em Proceedings of the National academy of Sciences of the United
  States of America} 106(38):16245--16250.

\bibitem{Bischofs2003}
Bischofs IB, Schwarz US (2003) Cell organization in soft media due to active
  mechanosensing.
\newblock {\em Proceedings of the National academy of Sciences of the United
  States of America} 100(16):9274--9297.

\bibitem{Zemel2010}
Zemel A, Rehfeldt F, Brown AEX, Discher DE, Safran SA (2010) Optimal matrix
  rigidity for stress-fibre polarization in stem cells.
\newblock {\em Nature Physics} 6(6):468--473.

\bibitem{Kabaso2011}
Kabaso D, Shlomovitz R, Schloen K, Stradal T, Gov NS (2011) Theoretical model
  for cellular shapes driven by protrusive and adhesive forces.
\newblock {\em PLoS Computational Biology} 7(5):e1001127.

\bibitem{Ramos2018}
Ramos JR, Travasso R, Carvalho J (2018) Capillary network formation from
  dispersed endothelial cells: Influence of cell traction, cell adhesion, and
  extracellular matrix rigidity.
\newblock {\em Physical Review E} 97(1):012408.

\bibitem{Copos2017}
Copos CA, et~al. (2017) Mechanosensitive adhesion explains stepping motility in
  amoeboid cells.
\newblock {\em Biophysical Journal} 112(12):2672--2682.

\bibitem{Uatay2018}
Uatay A (2018) A stochastic modeling framework for single cell migration:
  coupling contractility and focal adhesions.
\newblock {\em arXiv:1810.11435. Preprint, posted Oct 26, 2018.}

\bibitem{Feng2018}
Feng J, Levine H, Mao X, Sander LM (2018) Stiffness sensing and cell motility:
  Durotaxis and contact guidance.
\newblock {\em bioRxiv Preprint posted May 11, 2018}.

\bibitem{Welf2012}
Welf ES, Naik UP, Ogunnaike BA (2012) A spatial model for integrin clustering
  as a result of feedback between integrin activation and integrin binding.
\newblock {\em Biophysical Journal} 103(6):1379--1389.

\bibitem{Humphries2007}
Humphries JD, et~al. (2007) Vinculin controls focal adhesion formation by
  direct interactions with talin and actin.
\newblock {\em Journal of Cell Biology} 179(5):1043--1057.

\bibitem{Sawada2006}
Sawada Y, et~al. (2006) Force sensing by mechanical extension of the src family
  kinase substrate p130cas.
\newblock {\em Cell} 127(5):1015--1026.

\bibitem{Bos2003}
Bos JL, et~al. (2003) The role of rap1 in integrin-mediated cell adhesion.
\newblock {\em Biochemical Society Transactions} 31(Pt 1):83--86.

\bibitem{Block2011}
Ali O, et~al. (2011) Cooperativity between integrin activation and mechanical
  stress leads to integrin clustering.
\newblock {\em Biophysical Journal} 100(11):2595--2604.

\bibitem{Graner:1992ve}
Graner F, Glazier JA (1992) Simulation of biological cell sorting using a
  two-dimensional extended potts model.
\newblock {\em Physical Review Letters} 69(13):2013--2016.

\bibitem{Albert2014}
Albert PJ, Schwarz US (2014) Dynamics of cell shape and forces on
  micropatterned substrates predicted by a cellular potts model.
\newblock {\em Biophysical Journal} 106(11):2340--2352.

\bibitem{Magno2015}
Magno R, Grieneisen VA, Mar{\'{e}}e AFM (2015) The biophysical nature of cells:
  Potential cell behaviours revealed by analytical and computational studies of
  cell surface mechanics.
\newblock {\em BMC Biophysics} 8(8).

\bibitem{Boettiger2007}
Boettiger D (2007) Quantitative measurements of integrin-mediated adhesion to
  extracellular matrix.
\newblock {\em Methods in Enzymology} 426:1--25.

\bibitem{Davies:2011wr}
Davies AJ (2011) {\em The Finite Element Method: An Introduction With Partial
  Differential Equations}.
\newblock (Oxford University Press, Oxford).

\bibitem{Changede2016}
Changede R, Sheetz M (2016) Prospects {\&} overviews integrin and cadherin
  clusters: A robust way to organize adhesions for cell mechanics.
\newblock {\em Bioessays} 93(1):1--12.

\bibitem{Ambrosi:2006tv}
Ambrosi D (2006) Cellular traction as an inverse problem.
\newblock {\em SIAM Journal on Applied Mathematics} 66(6):2049--2060.

\bibitem{Knorr2012}
Knorr M, Koch D, Fuhs T, Behn U, Kas JA (2011) Stochastic actin dynamics in
  lamellipodia reveal parameter space for cell type classification.
\newblock {\em Soft Matter} 7:3192--3203.

\end{thebibliography}

\newpage

\section*{Supporting Material}

\subsection*{Supplementary methods}

In the main text, we proposed that matrix stress induces focal adhesion strengthening but noted that matrix stress might also reinforce cell contractility. Supplementary Figure{7} shows the results of having $\vec{F}_s = \vec{F}_s \cdot \left ( 1+p \frac{g(\dund{\sigma}(\vec{x}'))}{\sigma_h+g(\dund{\sigma}(\vec{x}'))} \right )$ instead of focal adhesion strengthening as described in the main text. Since matrix stresses are defined on the lattice sites while forces are defined on the nodes of the lattice, we needed to assume some interpolation. We choose to take

\begin{equation}
\vec{F}_s = \vec{F}_s \cdot \frac{1}{4} \sum_{\mathrm{surrounding 4 nodes}}\left ( 1+p \frac{g(\dund{\sigma}(\vec{x}'))}{\sigma_h+g(\dund{\sigma}(\vec{x}'))} \right ).
\end{equation}

\subsection*{Supplementary videos}

\begin{enumerate}[leftmargin=2cm]
\item[\textbf{Video S1}] Cell spreading on substrates of 1,5 and 50kPa (Model 1). \\ These are time series of Figure 2A of 500 MCS,.
\item[\textbf{Video S2}] Cell spreading on substrates of 1,5 and 50kPa (Model 2-version1). \\ These are time series of Figure 3A of 500 MCS,.
\item[\textbf{Video S3}] Cell spreading on substrates of 1,5 and 50kPa (Model 2-version2). \\ These are time series of Figure S7A of 500 MCS.
\item[\textbf{Video S4}] Cell durotacting on substrate with rigidity gradient 20 Pa/$\upmu$m .
\end{enumerate}

\begin{table}[H]
\centering

\begin{tabular}{|p{1.5cm}|p{3cm}|p{1.5cm}|p{2.5cm}|p{3cm}|}
\hline
parameter & description & value & unit & value was \\ \hline \hline
\textbf{CPM} & & &  & \\ \hline \hline
$\Delta x$ & lattice site width& 2.5 & $\upmu$m & chosen \\ \hline 
$\lambda$ & area constraint/cell stiffness & 0.0002 & N/m per lattice site$^2$ & chosen \\ \hline
$J(0,\mathrm{cell})$ &adhesive energy& 3000 & Nm per lattice site & chosen \\ \hline
$nbo$ & neighbourhood radius for adhesive energy & 10 & - & estimated based on accuracy of line tension \cite{Magno2015} \\ \hline
$\lambda_C$ & adhesion strength &600 & Nm per lattice site  & chosen \\ \hline
$A_h$ & area saturation & 1000 & lattice sites & chosen \\ \hline
$\lambda_N$ & focal adhesion strength & 4 & Nm & chosen \\ \hline
$p$ & actin-integrin strength & 1 & - & chosen \\  \hline
$\sigma_h$ & saturation actin-integrin binding & 5000 & N/$m^2$ & chosen \\ \hline
$T$ & cellular temperature & 2 & Nm & chosen \\ \hline
\hline
\textbf{Forces}& & &  & \\ \hline \hline
$\mu$ &traction magnitude & 0.001 & Nm per lattice site & estimated based on endothelial traction stresses \cite{Califano2010} \\ \hline
$v_0$ & free velocity of myosin molecules & 100 & nm/s & estimated based on non-muscle myosin IIB \cite{Norstrom2010,Vogel2013} \\ \hline
$E$ & Young's modulus& 10000 & N/$m^2$ & varies \\ \hline
$\nu$ & Poisson's ratio& 0.45 & - & chosen \\ \hline 
$\tau$ & substrate thickness &10  & $\upmu$m & \cite{Ambrosi:2006tv} \\ \hline
\hline
\textbf{FA's} & & &  & \\ \hline \hline
$\gamma$ & growth rate & 0.05 & /s & estimated \cite{Novikova2013} \\ \hline
$N_0$ & size initial adhesion &5000 & - & estimated based on nascent adhesions \cite{Changede2016} \\ \hline
$N_m$ & maximum free bonds &8000000 & - & chosen \\ \hline
$N_b$ & maximum size focal adhesion&39062 & - & estimated based on number of integrins that fit in one lattice site \cite{Changede2016} \\ \hline
$\phi_s$ & slip tension & 4.02 & pN/$m^2$ & \cite{Novikova2013} \\ \hline
$\phi_c$ & catch tension  & 7.76 & pN/$m^2$ & \cite{Novikova2013} \\ \hline
$t_\mathrm{FA}$ & focal adhesion growth time & 10 & s & estimated based on protrusion lifetimes \cite{Knorr2012} \\ \hline
$\Delta t_\mathrm{FA}$ & time steps& 0.01 & s& chosen \\  \hline
\end{tabular}

\caption*{Table S1. Parameter values.}
\label{tab:par}
\end{table}

\begin{figure}
\centering
\includegraphics[width=\textwidth]{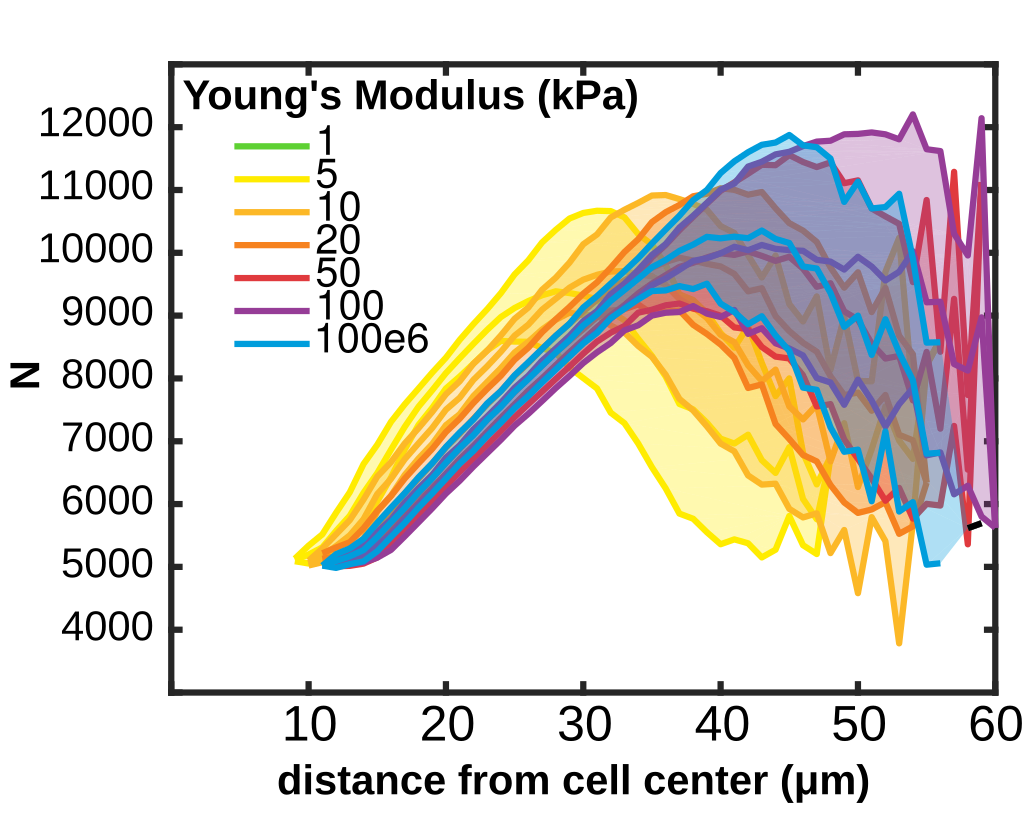}
\caption*{Figure S1. The number of integrin bonds per cluster ($N$) in model M1 as a function of distance from the cell center. All clusters at 2000 MCS from 25 simulations were pooled. Shaded regions show standard deviations.  }
\end{figure}

\begin{figure}
\centering
\includegraphics[width=\textwidth]{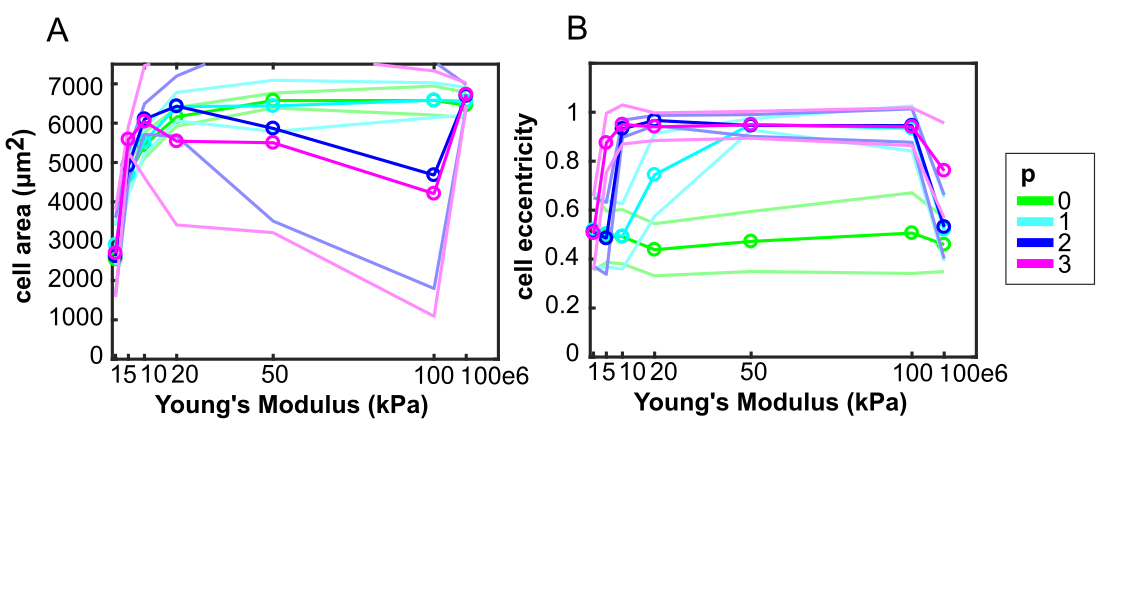}
\caption*{Figure S2. Model sensitivity to actin-integrin strength $p$. (A) Cell area as a function of substrate stiffness, shaded regions: standard deviations over 25 simulations;  (B) Cell eccentricity as a function of substrate stiffness, shaded regions: standard deviations over 25 simulations. }
\end{figure}

\begin{figure}
\centering
\includegraphics[width=\textwidth]{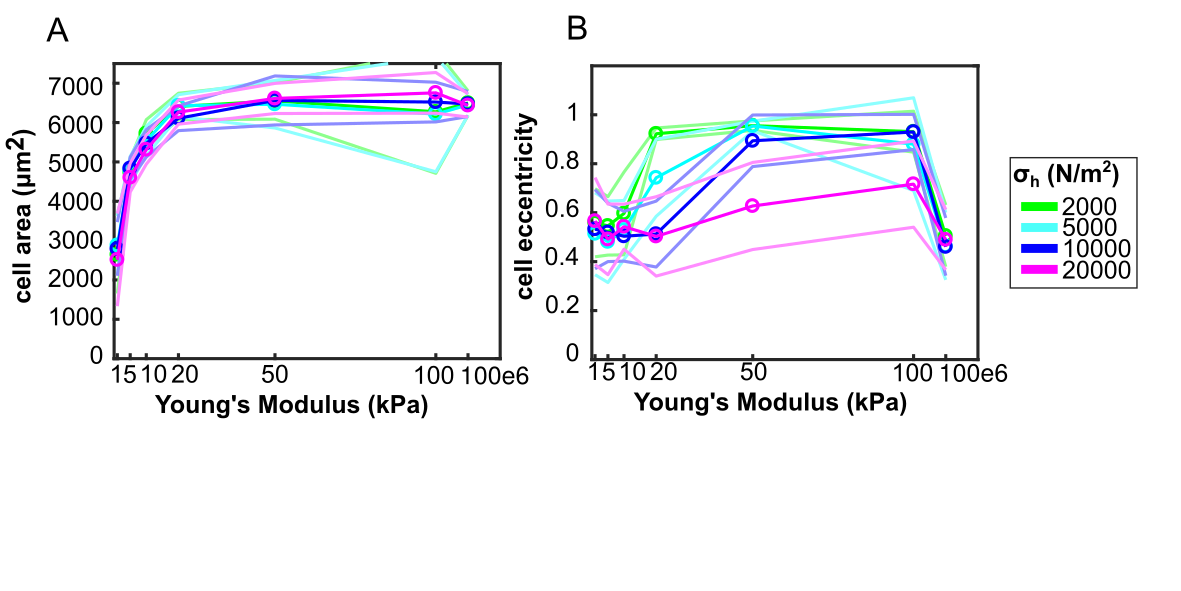}
\caption*{Figure S3. Model sensitivity to saturation value for actin-integrin strength $\sigma_h$. (A) Cell area as a function of substrate stiffness, shaded regions: standard deviations over 25 simulations;  (B) Cell eccentricity as a function of substrate stiffness, shaded regions: standard deviations over 25 simulations.}
\end{figure}

\begin{figure}
\centering
\includegraphics[width=\textwidth]{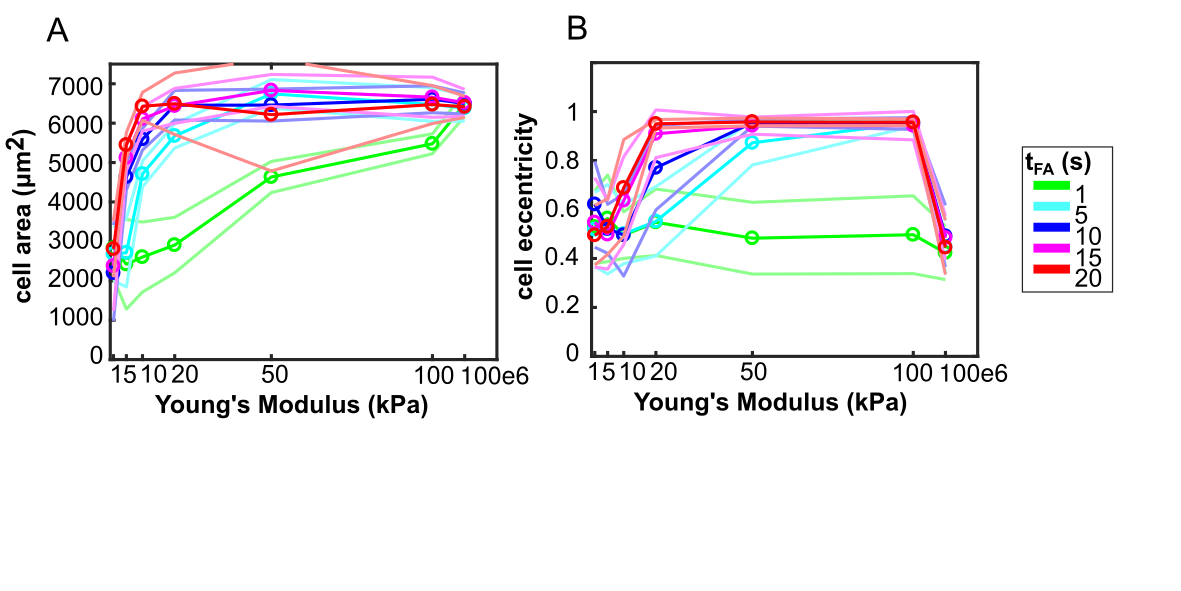}
\caption*{Figure S4. Model sensitivity to focal adhesion growth time $t_{\mathrm{FA}}$. (A) Cell area as a function of substrate stiffness, shaded regions: standard deviations over 25 simulations;  (B) Cell eccentricity as a function of substrate stiffness, shaded regions: standard deviations over 25 simulations. }
\end{figure}

\begin{figure}
\centering
\includegraphics[width=\textwidth]{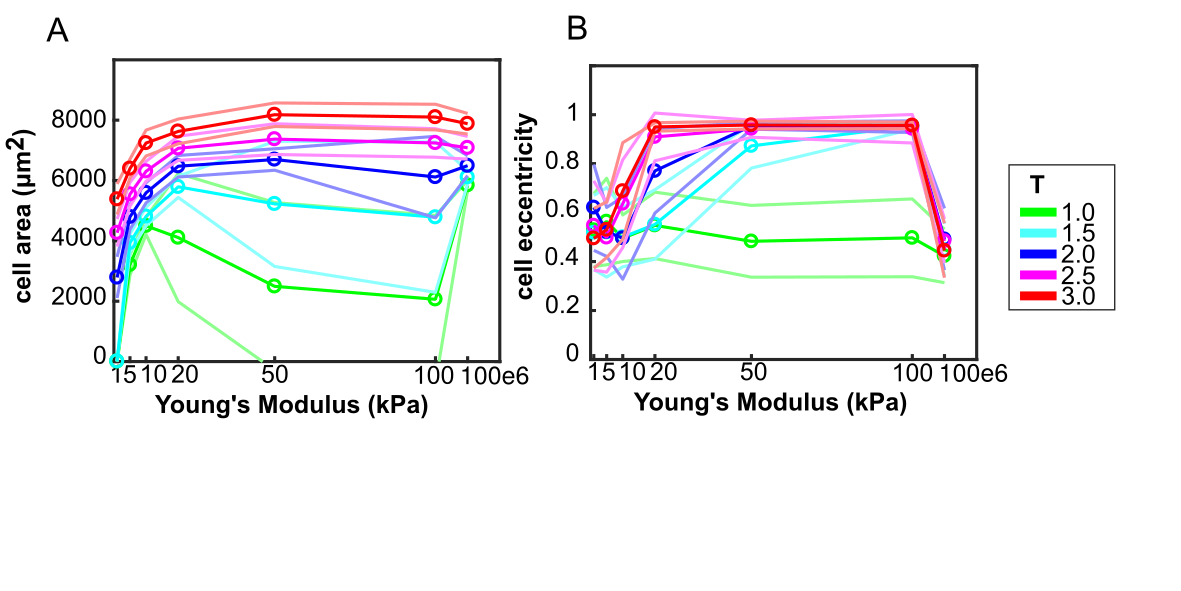}
\caption*{Figure S5. Model sensitivity to cellular temperature $T$. (A) Cell area as a function of substrate stiffness, shaded regions: standard deviations over 25 simulations;  (B) Cell eccentricity as a function of substrate stiffness, shaded regions: standard deviations over 25 simulations. }
\end{figure}

\begin{figure}
\centering
\includegraphics[width=\textwidth]{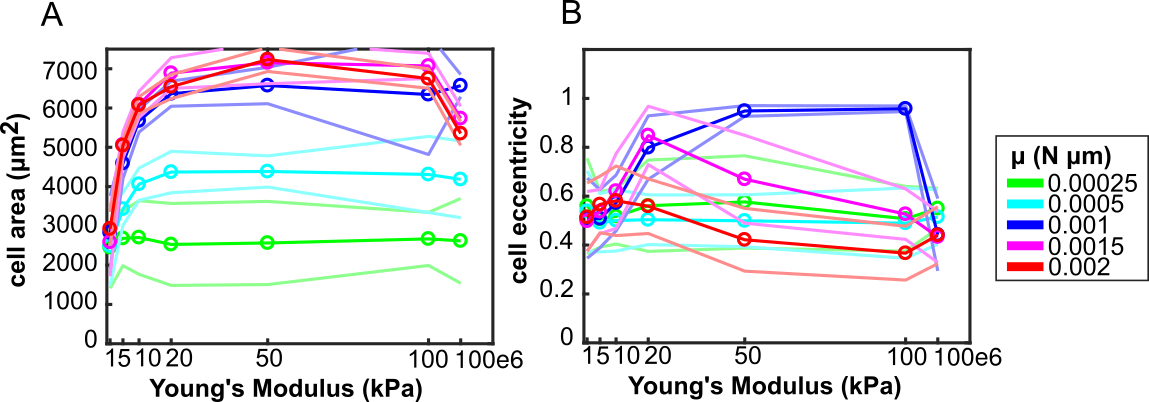}
\caption*{Figure S6. Model sensitivity to traction force magnitude $\mu$. (A) Cell area as a function of substrate stiffness, shaded regions: standard deviations over 25 simulations;  (B) Cell eccentricity as a function of substrate stiffness, shaded regions: standard deviations over 25 simulations.}
\end{figure}

\begin{figure}
\centering
\includegraphics[width=\textwidth]{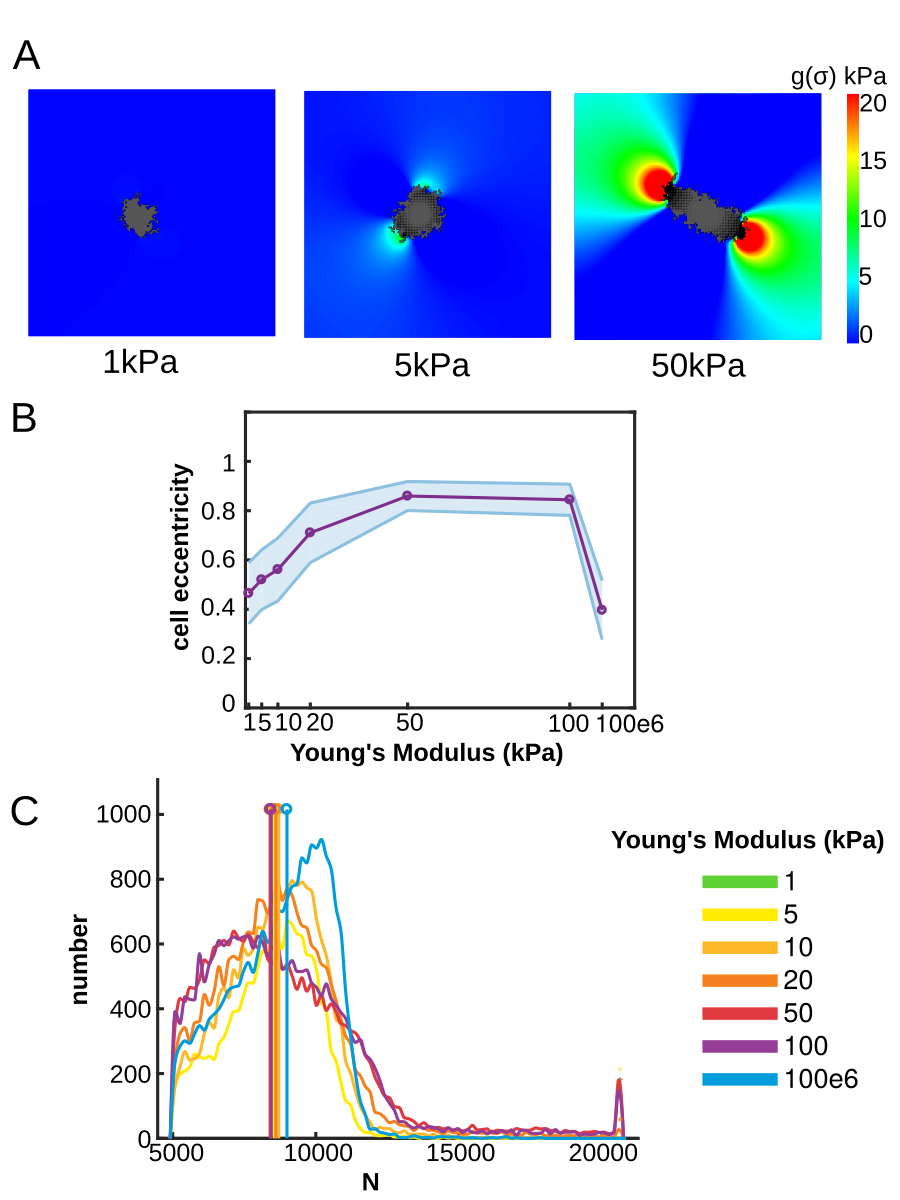}
\caption*{Figure S7. Cells elongate on substrates of intermediate stiffness. Model M2 was used where matrix stress reinforces traction force $\vec{F}_s$ with $p=5$. (A) Example configurations of cells at 2000 MCS on substrates of 1,50 and 50 kPa. Colors: hydrostatic stress; (B) Cell eccentricity as a function of substrate stiffness, shaded regions: standard deviations over 25 simulations; (C) distribution of N, the number of integrin bonds per cluster, all focal adhesion at 2000 MCS from 25 simulations were pooled. We indicate the median. Color coding (C): See legend next to (C).}
\end{figure}

\begin{figure}
\centering
\includegraphics[width=\textwidth]{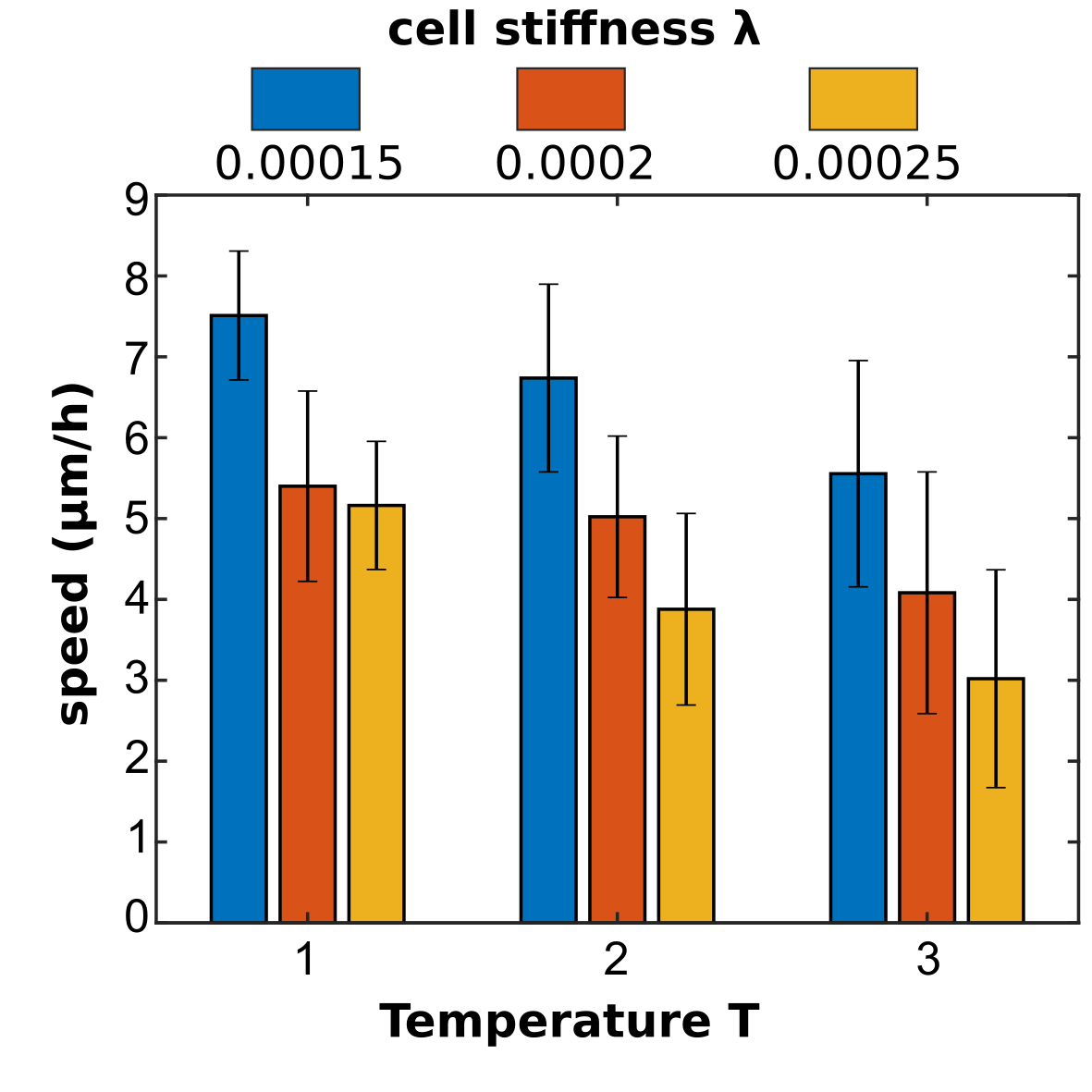}
\caption*{Figure S8. Durotaxis speed in $\upmu$m/h as a function of cell stiffness $\lambda$ and cellular temperature $T$. Values: mean $\pm$ standard deviation over 25 simulations. }
\end{figure}

\end{document}